\documentclass[aps,prb,twocolumn,groupedaddress,longbibliography]{revtex4-2}

\usepackage{amsmath}
\usepackage{amsfonts}
\usepackage{amssymb}
\usepackage{subcaption}
\usepackage{graphicx}
\usepackage{verbatim}
\DeclareGraphicsExtensions{.jpg,.png,.jpeg,.pdf,.eps}
\usepackage{physics}
\usepackage{xcolor}
\usepackage{enumerate}
\usepackage{hyperref}
\newcommand{\I}{\mathrm{i}}
\newcommand{\E}{\mathrm{e}}

\begin{document}

\title{{{Identification} of the transport regimes of the one-dimensional Holstein model}}

\author{Suzana Miladi\'c, Nenad Vukmirovi\'c}
\email{nenad.vukmirovic@ipb.ac.rs}

\affiliation{Institute of Physics Belgrade, University of Belgrade,  Pregrevica 118, 11080 Belgrade, Serbia}

\date{\today}

\begin{abstract}
The Holstein model is a benchmark model of systems with electron-phonon interaction. However, its electrical transport properties are not yet fully understood. In this work, we performed numerically exact calculations of imaginary-time current-current correlation function of the Holstein Hamiltonian for a broad range of model parameters. These calculations were performed using a path-integral based Quantum Monte Carlo method. We compared these results with the results obtained under the assumption of conventional band transport, small polaron hopping and polaron band transport. From this comparison we identified the regions in parameter space where each of these transport regimes is valid. In some cases when imaginary-time comparison of current-current correlation functions could not give conclusive results, we complemented them with real-time comparisons or the comparison with literature data on numerically exact dc mobilities. Overall, we found that the parameter space is almost completely covered by the three mentioned transport regimes.
\end{abstract}

\maketitle

\section{Introduction}\label{sec:1}
The mobility of charge carriers in semiconductor materials is mainly governed by the interaction between electrons and phonons \cite{Jacoboni, Ashcroft, Mahan}. Depending on the strength of the interaction and the temperature, different carrier transport regimes can occur \cite{JPSJ.33.327, EminD}. In scenarios where the electron-phonon interaction is weak, carriers are delocalized and sometimes scatter on the phonons, placing the system within the traditional band transport regime. This is the common regime observed in the conduction band of traditional semiconductors, where transport phenomena are described by the Boltzmann equation \cite{prb104-085203,rpp83-036501,prb97-121201,prb97-045201,prb95-075206,prb94-201201,prb94-085204}. In conditions of strong electron-phonon interaction and high temperatures, the carriers become entirely localized, and they move by occasionally hopping between neighboring sites \cite{ap21-494,ap24-305,prb4-3639,prb56-4484,jssc12-246,jpcs28-581}. This type of hopping behavior commonly occurs in small molecule based organic semiconductors \cite{prb43-11720}. The hopping rates and mobility in these systems are often modeled using the Marcus formula \cite{cm20-3205,nc2-437,jpcb113-8813,pccp12-11103,cr107-926}. In cases where the interaction remains strong while the temperature is low, carriers delocalize and form a band, which is much narrower than the bare electron band. This scenario is known as the polaron band transport regime \cite{pr135-A233}.

The mentioned regimes take place under conditions of either a weak or strong electron-phonon interaction, or in scenarios where the temperature is significantly high or low. The methods that have been developed to calculate the carrier mobility in real materials, are typically based on the assumption of either band transport \cite{prb104-085203,rpp83-036501,prb97-121201,prb97-045201,prb95-075206,prb94-201201,prb94-085204} or hopping mechanisms \cite{cm20-3205,nc2-437,jpcb113-8813,pccp12-11103,cr107-926}.
 However, these methods are sometimes applied even when it is uncertain if one of these regimes is genuinely present within the material. On the other hand, the methods that do not make the assumption about the transport regimes are typically limited to model Hamiltonians \cite{prb79-235206,njp12-023011,jcp128-114713,jpcl4-1888,prl96-086601,pccp17-12395,pccp18-1386,prb85-245206,ap391-183,jpcl5-1335,prb69-075211,prl129-096401,prb107-125165,prb112-035111,prb112-035112,prl114-146401,prr1-033138,jpcb115-5312,prl91-256403,ncomms12-4260,cphys6-125,prb111-195140}.
 For these reasons, it is of significant interest to understand the range of validity of different transport regimes. Moreover, it is especially important to determine if the specified regimes cover the entire parameter space or if there are areas where carrier transport cannot be adequately explained by any of these regimes.

This research tackles these questions by examining a prototype model with electron-phonon interaction, the Holstein model \cite{ap8-343}. An ideal approach to answering these questions would be to perform numerically exact calculations of carrier mobility or current-current correlation function using real-time or real-frequency methods throughout the whole parameter space and compare the results with the predictions of each of the theories corresponding to characteristic transport regimes. Although several numerically exact approaches have been applied to the Holstein model in recent years, their practical applicability is restricted to a limited part of the parameter space \cite{prb100-094307,prb102-165155,jpcl11-4930,jcp153-020901,prb105-054311,jcp159-094113}. On the other hand, imaginary-time quantities can be reliably calculated across a significantly broader range of the parameter space.

In this work, we therefore perform numerically exact calculations of the imaginary-time current-current correlation function over a broad range of parameters. These calculations are conducted employing a path-integral based Quantum Monte Carlo (QMC) method \cite{pr97-660,pr127-1004,prb27-6097,prb30-1671,prb107-184315}. For the same parameter sets, we calculate the same quantity under the assumptions of band transport, hopping, or polaron band transport mechanisms. The comparison of these results allows us to infer the range of validity of each of the transport regimes. In some cases where the comparisons of the imaginary-time correlation functions do not give conclusive results, we complement them with comparisons of real-time correlation functions or mobilities.

The structure of the paper is as follows. Section \ref{sec:2} presents the model and methods on which our work is based. In Sec. \ref{sec:2a}, we briefly introduce the Holstein Hamiltonian. Section \ref{sec:2b} outlines the QMC method employed to compute current-current correlation functions and dc mobilities, which are pivotal to our findings. In Sec. \ref{sec:2c}, we derive analytical expressions for various transport regimes and present some results based on these expressions. Specifically, Sec. \ref{sec:2c1} derives formulas for current-current correlation functions and dc mobility for strong electron-phonon interactions, with a particular emphasis on the high temperature limit. Next, Sec. \ref{sec:2c3} derives expressions in the case of weak electron-phonon interaction and large polaron transport. Lastly, Sec. \ref{sec:2c2} addresses low-temperature polaron narrow band transport.  In Sec.~\ref{sec:2c4} we present some results obtained from the formulas presented in Secs.~\ref{sec:2c1}-\ref{sec:2c2}. In Sec. \ref{sec:3}, we compare the results obtained using the expressions  for different transport regimes to numerically exact QMC results and construct the transport regime diagram which gives the information about the range of parameters where each of the transport regimes takes place. We finalize the paper with the conclusions in Sec.~\ref{sec:4}.

\section{Model and methods}\label{sec:2} 

\subsection{Holstein Hamiltonian}\label{sec:2a}

{We consider the model in one dimension consisting of one carrier (electron) interacting with a chain of equidistant motifs (ions). This simple model is described with one-dimensional Holstein Hamiltonian}
\begin{equation}
\begin{split}
H=&-J\sum_{p}\sum_{\gamma=\pm 1}a_{p+\gamma}^{\dagger}a_p + \sum_p \omega_0b_p^{\dagger}b_p+ \\
  &\sum_p G \: a_p^{\dagger}a_p \left(b_p+b_p^{\dagger}\right),
\end{split}
\end{equation}
where $a_p$ and $a_p^{\dagger}$ are electronic operators that annihilate or create an electron at lattice site $p$, $b_p$ and $b_p^{\dagger}$ are phonon operators that annihilate or create a phonon of angular frequency $\omega_0$ at lattice site $p$, $G$ is the electron-phonon coupling constant and $J$ is the electronic transfer integral. We use the system of units where the reduced Planck constant $\hbar$, the elementary charge $e$, the lattice constant $a_l$ and the Boltzmann constant $k_B$ are set to 1.

\subsection{QMC method}\label{sec:2b}
The main quantity that we evaluate in this work is the current-current correlation function which contains information about electronic transport properties. It reads
\begin{equation}
    \label{path_int}
    C_{jj}\qty(t)=\langle j(t)j(0)\rangle = Z^{-1}\mathrm{Tr}\left[\E^{-\beta H}\E^{\I tH}j\E^{-\I tH}j \right]\phantom{.},
\end{equation}
where $Z=\mathrm{Tr}\E^{-\beta H}$ represents the partition function in the canonical ensemble, $\beta$ is the inverse temperature and the operator $j$ is given by
\begin{equation}
    j=\I J \sum\limits_p \left(a_p^{\dagger}a_{p+1}-a_{p+1}^{\dagger}a_p \right)\phantom{.}.
\end{equation}
The operator $j$ can also be represented as
\begin{equation}\label{eq:jk}
j=\sum_k v_k a_k^\dagger a_k,
\end{equation}
where $v_k=-2J\sin k$ is the band velocity and the operators $a_k$ ($a_k^\dagger$) annihilate (create) an electron of momentum $k$.
When $t$ is a real number, Eq.~\eqref{path_int} defines the real-time current-current correlation function. For $t=-\I\tau$, where the real number $\tau$ is from the interval $0\le \tau\le \beta$, Eq.~\eqref{path_int} defines the imaginary-time current-current correlation function.
The dc mobility is related to the real-time current-current correlation function as \cite{Mahan}:
\begin{equation}
    \label{kubo_formula}
    \mu=\dfrac{\beta}{2}\int\limits_{-\infty}^{\infty}\mathrm{d}t\langle j(t)j(0)\rangle \phantom{.}.
\end{equation}
{The system with a single electron in the canonical ensemble is equivalent to the system in the grand canonical ensemble with chemical potential $\mu\to -\infty$. This corresponds to the physical situation where the carrier concentration in the band is low and the chemical potential is well below the bottom of the band.}

We perform numerically exact calculation of the current-current correlation function using the path-integral Quantum Monte Carlo method. The details of the methodology were presented in our previous work \cite{prb107-184315} and summarized in Ref. \cite{prb109-214312}, so we discuss it here only briefly. Within the method, the real- or imaginary-time evolution operators ($\E^{\pm\I Ht}$ or $\E^{-\beta H}$) are divided into evolution operators over small time steps using the Suzuki-Trotter expansion. The path integral for the correlation function then reduces to a sum which is evaluated using a Monte Carlo procedure. The systematic error arising from finite time step is controlled by choosing sufficiently small time step. The statistical error coming from a finite number of samples in Monte Carlo evaluation of the sum is controlled by choosing sufficiently large number of samples. It is therefore more challenging to obtain results at extremely low temperatures (large $\beta$) or at greater real-times (large $t$).

To perform more efficient calculations, it is also possible to exploit the fact that the trace in Eq. \eqref{path_int} can be expressed using any complete basis that spans the Hilbert space of the system, as we discussed in our previous work \cite{prb107-184315}. While the final result is independent of the choice of the basis, the basis does influence the statistical error in the Monte Carlo summation that is performed over a limited number of terms only \cite{sa6-eabb8341}. This insight enables us to select a basis that reduces the statistical error of Monte Carlo sums within certain parameter ranges. With such a choice, we can perform the simulation for longer real times before the dynamical sign problem \cite{pre61-5961,jcp110-12,pra41-5709,jcp108-3871,cpc63-415,prl115-266802} occurs and we can perform imaginary-time calculations at lower temperatures (larger $\beta$). We find that it is convenient to use the basis of electron momentum states and the coordinate representation for phonons in case of weaker electron-phonon interactions. For stronger interaction we find that it is more convenient to use the position representation for the electron and the coordinate representation for phonons. This basis gives an additional advantage that phonon coordinates can be integrated out.

{In this paragraph we {give some} technical details regarding our QMC calculations. We have {performed calculations so} that the standard deviation is no greater than about 1\% of the result. To achieve such precision we used from $10^3$ to $10^6$ MC samples in a single calculation. Full QMC calculations were repeated 10 or 100 times for better statistics and standard error estimation. We have {chosen} the time discretization step values to be less or equal to $0.1/J$ since convergence tests showed us that for such small values there is no significant change in the result.
{Finally, we discuss} the system size parameter - the number of sites in the one-dimensional (1D) model. All our calculations were {performed} to be representative of the thermodynamic limit.  We have chosen {sufficiently large} number of sites so that its futher increase does not affect the {result}.
For the majority of the calculations for short real times ($Jt\leq 1.0$) and imaginary times (at higher temperatures) the number of about 10 sites was sufficient, while for longer real times and some lower temperatures there was a need for a much larger system with {several} tens of sites.  This also sets the limit to our calculations {as these are not computationally feasible for larger systems}. {In Fig. S9 of Supplemental Material \cite{supplement} we present an example of the calculations for different system size when $\lambda=0.01$, $T/J=0.1$ and $\omega_0=3J$. We also {present in Fig. S10 of Supplemental Material \cite{supplement}}  a detailed table with the parameter values (time steps, number of sites, number of MC samples,...).}
We refer the reader to Ref.~\cite{prb107-184315} for more detailed discussion regarding the choice of simulation parameters.}

\subsection{Transport regimes}\label{sec:2c}

In the case of 1D Holstein model, the strong electron-phonon coupling causes the carrier to self-trap in the potential well which is formed due to phonon displacements on the lattice site \cite{EminD}. In such a way, so called "localized small polaron" is formed. At very low, near zero, temperatures, a small polaron can tunnel between equal-energy neighboring sites with no change in energy (diagonal/coherent processes). This is similar to free carrier band transport, but in this case the polaron band is extremely narrow \cite{pr135-A233}. At higher temperatures, the polaron may move by phonon-assisted hopping between sites (non-diagonal/incoherent processes). On the other hand, in the case of weak electron-phonon coupling, the carrier is usually called a large polaron. It acts as a free carrier with somewhat increased effective mass that  occasionally scatters on lattice vibrations. The phonon scattering of large polaron causes its mobility to decrease with increasing temperature and its transport is conventional band transport. The large polaron is in general more mobile compared to the small polaron. In this section we present the expressions for current-current correlation function in each of the three mentioned regimes: (small polaron) hopping, (small) polaron band transport and (large polaron) band transport.

\subsubsection{Hopping transport}\label{sec:2c1}

The expression for hopping rate between two neighboring sites can be obtained by treating the electronic coupling between the sites as a perturbation and applying the Fermi's golden rule. {Such an expression is expected to be valid when the electron-phonon interaction is sufficiently strong and when the temperature is not too low.}
This expression takes the form {which is known in the literature}\cite{acpk, tomic2017, prb79-115203}
\begin{equation}\label{eq:whop}
W=J^2\int_{-\infty}^{\infty}\dd t \: \E^{-2\qty(\frac{G}{\omega_0})^2
\qty[2n_{\mathrm{ph}}+1-\qty(n_{\mathrm{ph}}+1)\E^{-\I\omega_0t}
-n_{\mathrm{ph}}\E^{\I\omega_0t}]},
\end{equation}
where $n_{\mathrm{ph}}=\qty(\E^{\beta\omega_0}-1)^{-1}$ is the phonon occupation number.
The derivation of this expression can be found in Ref. \cite{acpk} or \cite{tomic2017}.
The corresponding current-current correlation function reads [see Eq. (24) in Ref.~\cite{prb109-214312}]
\begin{equation}\label{eq:jj0hop}
C_{jj}\qty(t)=2J^2 \: \E^{-2\qty(\frac{G}{\omega_0})^2
\qty[2n_{\mathrm{ph}}+1-\qty(n_{\mathrm{ph}}+1)\E^{-\I\omega_0t}
-n_{\mathrm{ph}}\E^{\I\omega_0t}]}.
\end{equation}
In the limit of high temperature Eq.~\eqref{eq:whop} reduces to the widely used Marcus formula \cite{cm20-3205,nc2-437,jpcb113-8813,pccp12-11103,cr107-926}
\begin{equation}
W=J^2\sqrt{\dfrac{\beta\pi}{\lambda}}\E^{-\frac{\beta\lambda}{4}} \phantom{.},
\end{equation}
with $\lambda=\frac{2G^2}{\omega_0}$ and the corresponding mobility
\begin{equation}\label{mu_slhtA}
\mu=\beta J^2\sqrt{\dfrac{\beta\pi}{\lambda}}\E^{-\frac{\beta\lambda}{4}} \phantom{.}.
\end{equation}
The corresponding current-current correlation function reads [see Eq. (26) in Ref.~\cite{prb109-214312}]
\begin{equation}
    \label{Cjj_slhtA}
    C_{jj}(t)=2J^2\E^{-\sigma_0^2t^2}\E^{-\I\beta\sigma_0^2t}\phantom{.},
\end{equation}
where $\sigma_0^2=\frac{2G^2}{\beta\omega_0}$.

In the case of Holstein model with a single dispersionless phonon mode that we are considering, Eq.~\eqref{eq:whop} leads to a diverging result for the hopping rate and the dc mobility since the current-current correlation function from Eq.~\eqref{eq:jj0hop} does not decay to zero when $t\to\infty$. The origin of this divergence comes from the fact that small polarons are treated as quasiparticles with infinite lifetime in such considerations. An improvement over such approach where broadening of small polaron states is taken into account by calculating the relevant self-energies was presented {recently} in Appendix C of Ref.~\cite{prb109-214312} {and here we present the first application of that approach}. One then obtains [see Eq. (25) in Ref.~\cite{prb109-214312}]
 \begin{equation}\label{Cjj_slA}
     C_{jj}(t)=g(t)
      \E^{-2\qty(\frac{G}{\omega_0})^2\qty[2n_{\mathrm{ph}}+1 -(n_{\mathrm{ph}} +1)\E^{-i\omega_0 t} - n_{\mathrm{ph}}\E^{\I\omega_0 t} ]}\phantom{.},
 \end{equation}
where
\begin{equation}
g(t)=2J^2\dfrac{\beta}{t(\beta-\I t)}\dfrac{1}{\sqrt{c_0}}\dfrac{I_1\left(-2(\beta-\I t)\sqrt{c_0} \right)J_1(2t\sqrt{c_0})}{I_1\qty(-2\beta\sqrt{c_0})}\phantom{.},
\end{equation}
while $I_1$ ($J_1$) is the (modified) Bessel function of the first kind of order 1 and
\begin{equation}
c_0=2J^2\E^{-2\qty(\frac{G}{\omega_0})^2\qty(2n_{\mathrm{ph}}+1)}I_0\qty(\alpha)
\end{equation}
with
\begin{equation}
\alpha=4\qty(\frac{g}{\omega_0})^2\sqrt{n_{\mathrm{ph}}\qty(n_{\mathrm{ph}}+1)}.
\end{equation}
The corresponding dc mobility, that is related to $C_{jj}(t)$ via Eq.~\eqref{kubo_formula}, is then given as
\begin{equation}\label{mu_slA}
\begin{split}
\mu =&
2J^2\beta\pi
\E^{-2\qty(\frac{G}{\omega_0})^2\qty(2n_{\mathrm{ph}}+1)} \times \\
& \frac{\sum_{l=-\infty}^{\infty}\int \dd \omega e^{-\beta\omega}A\qty(\omega) A\qty(\omega+l\omega_0)I_l\qty(\alpha) e^{-\frac{l\omega_0\beta}{2}}}
{\int \dd \omega e^{-\beta\omega}A\qty(\omega)},
\end{split}
\end{equation}
where
\begin{equation}
 A\qty(\omega)=\frac{\sqrt{4c_0-\omega^2}}{2\pi c_0}\theta\qty(4c_0-\omega^2),
\end{equation}
$I_l$ denotes the modified Bessel function of the first kind of order $l$ and $\theta$ is the step function.

To summarize this section, in the rest of the paper we will use Eqs.~\eqref{Cjj_slA} and \eqref{mu_slA} as the result for the current-current correlation function and the dc mobility in the hopping regime. We will also compare these results to their high-temperature limits given by Eqs.~\eqref{Cjj_slhtA} and \eqref{mu_slhtA}. The results for imaginary-time current-current correlation functions will be simply obtained by the replacement $t\to -\I\tau$.

\subsubsection{Band transport}\label{sec:2c3}

In this section, we present the expressions for the current-current correlation function and the dc mobility in the band transport regime. {This regime is expected to take place when the electron-phonon interaction is sufficiently weak and when the temperature is not too high.} Starting from Eq.~\eqref{eq:jk} we obtain
\begin{equation}
C_{jj}(t)=\frac{\sum_{k,k'}v_kv_{k'}\expval{a_k^\dagger(t)a_k(t)a_{k'}^\dagger a_{k'}}}{{\sum_k \expval{a_k^\dagger a_k}}}.
\end{equation}
We then make use of the independent particle (also known as bubble) approximation, that is known to be exact in the limit of weak electron-phonon interaction \cite{prb109-214312} for Holstein model:
\begin{equation}
\expval{a_k^\dagger(t)a_k(t)a_{k'}^\dagger a_{k'}}=
\delta_{kk'}\expval{a_k^{\dagger}(t)a_k}\expval{a_k(t)a_k^{\dagger}}
\end{equation}
and obtain
\begin{equation}
    \label{Cjj_wl_1a}
    {C_{jj}(t)}=4J^2\dfrac{\sum_k \langle a_k^{\dagger}(t)a_k\rangle\langle a_k(t)a_k^{\dagger}\rangle \sin^2(k)}{\sum_k \expval{a_k^\dagger a_k}} \phantom{.}.
\end{equation}
The expectation values in Eq.~\eqref{Cjj_wl_1a} are found from the electronic spectral function which is related to the retarded Green's function as $A_k(\omega)=-\frac{1}{\pi}\Im G_k^R(\omega)$. The Green's function is found from the self-energy in Migdal approximation which includes the diagrams that describe single phonon emission and absorption processes. This is in most cases sufficient for weak electron-phonon interaction. However, when $\frac{\omega_0}{2J}>1$ there are momenta $k$ that satisfy $\vert \varepsilon_k \pm \omega_0\vert > 2J$ for which single phonon processes are not allowed due to energy conservation. In these cases we also include the diagrams that describe two-phonon processes. The detailed expressions for self-energies and for expectation values in Eq.~\eqref{Cjj_wl_1a} are given in Sec. S1 of the Supplemental Material \cite{supplement}.

The dc mobility in the band transport regime is obtained from the expression (see Ref. \cite{prb99-104304}):
\begin{equation}
    \label{mu_wl0}
    \mu_{\mathrm{dc}} = \beta \dfrac{\sum_k n_k \tau_k v_k^2}{\sum_k n_k}\phantom{.},
\end{equation}
where $n_k=\E^{-\beta\varepsilon_k}$ and the scattering time is given as $\tau_k^{-1}=-2\Im \Sigma_k(\omega)\vert_{\omega=\varepsilon_k}$. {While the expression for band transport mobility including single phonon processes only has been previously used (see, e.g. \cite{prb99-104304}), we are not aware of previous works that include two phonon processes which are necessary when  $\frac{\omega_0}{2J}>1$.}

\subsubsection{Polaron band transport}\label{sec:2c2}

In this section, we derive the expression for the dc mobility and for the imaginary-time current-current correlation function in the polaron band transport regime. {This expression is expected to be valid when the electron-phonon coupling is strong and when the temperature is very low.}
To this end, we perform the Lang-Firsov unitary transformation of the Hamiltonian and work in the basis of polarons that form a narrow band. We treat the remaining interacting part of the Hamiltonian as a perturbation and evaluate the corresponding self-energy. We derive the expression for the current-current correlation function and identify the contributions stemming from: (i) processes with no phonon exchange (diagonal/coherent processes); (ii) phonon assisted processes (nondiagonal/incoherent processes). Since the processes (i) are the dominant ones in the polaron band transport regime, we include the contribution from these processes only in the expressions  for the polaron band transport regime.

First, we perform the Lang-Firsov \cite{jetp16-1301} unitary transformation
$\tilde{H}=U^{-1}HU$
with
\begin{equation}
U=e^{\sum_p \frac{G}{\omega_0} a_p^\dagger a_p \qty(b_p-b_p^\dagger)}.
\end{equation}
The transformed Hamiltonian takes the form $\tilde{H}=\tilde{H}_0+\tilde{V}$ with
\begin{equation}
 \tilde{H}_0 = \sum_k \tilde{\varepsilon}_k a_k^\dagger a_k + \sum_p \omega_0 b_p^\dagger b_p
\end{equation}
where
\begin{equation}
 \tilde{\varepsilon}_k = -\frac{G^2}{\omega_0} - 2 \tilde{J} \cos k
\end{equation}
is the polaron band dispersion and
\begin{equation}
 \tilde{J}=J \E^{-(2n_{\mathrm{ph}}+1)\frac{G^2}{\omega_0^2}}
\end{equation}
is the renormalized electronic transfer integral. The interaction term $\tilde{V}$ in the transformed Hamiltonian reads
\begin{equation}
 \tilde{V}=\frac{1}{N}\sum_{k,q} a_{k+q}^\dagger a_k \mathcal{B}_{k,q}
\end{equation}
with
\begin{equation}
 \mathcal{B}_{k,q}=-\sum_{p,r=p\pm 1} J
 \E^{\I\qty[\qty(k+q)p-kr]}
 \qty[\theta_p^\dagger \theta_r - \theta_0],
\end{equation}
where
$ \theta_p=\E^{\frac{G}{\omega_0}\qty(b_p-b_p^\dagger)}$
and
$ \theta_0=\E^{-(2n_{\mathrm{ph}}+1)\frac{G^2}{\omega_0^2}}$, while $N$ is the number of lattice sites.

The current-current correlation function in the independent particle (bubble) approximation~\cite{prb109-214312} is given as (special case of the result from Ref.~\cite{prb99-104304} when the Lang-Firsov transformation is used instead of a more general transformation)
\begin{equation}
 C_{jj}\qty(t)=\frac{\sum_{k_1k_2}
 \expval{a_{k_1}^\dagger\qty(t)a_{k_1}}
 \expval{a_{k_2}\qty(t)a_{k_2}^\dagger}
 Y_{k_1k_2}\qty(t)}{\sum_k \expval{a_{k}^\dagger a_k}}.
\end{equation}
The averages in the previous equation are taken with respect to the transformed Hamiltonian $\tilde{H}$, while
\begin{equation}\label{eq:yk1k2}
    Y_{k_1 k_2}(t)=-\dfrac{\tilde{J}^2}{N}\sum\limits_{\substack{X, Y=\pm 1, \\ Z}} XY\E^{\I k_1(X+Z)}\E^{\I k_2 (X-Z)} \theta_{XYZ}(t)\phantom{.}
\end{equation}
with
\begin{equation}
 \theta_{XYZ}\qty(t)=\E^{
 a_{0,Z,X,Z+Y}\qty[\qty(n_{\mathrm{ph}}+1) \E^{-\I\omega_0t}+
         n_{\mathrm{ph}} \E^{\I\omega_0t} ]}
\end{equation}
and
$a_{R_1R_2S_1S_2}=\frac{G^2}{\omega_0^2}(\delta_{R_1R_2}+\delta_{S_1S_2}-\delta_{R_1S_2}-\delta_{R_2S_1})$. We then divide
  $\theta_{XYZ}(t)=\theta_{XYZ}^{(0)} + \theta_{XYZ}^{(1)}(t)$, with the two components defined as follows:
\begin{equation}
    \begin{split}
        \theta_{XYZ}^{(0)}&=I_0\left[ 2a_{0,Z,X,Z+Y}\sqrt{n_{ph}(n_{ph}+1)}\right] \phantom{.}, \\
        \theta_{XYZ}^{(1)}(t)&=\sum\limits_{l\neq 0}I_l\left[2a_{0,Z,X,Z+Y}\sqrt{n_{ph}(n_{ph}+1)} \right] \\ & \times \E^{\I l\omega_0 t}\E^{-\frac{1}{2}l \omega_0 \beta} \phantom{.},
    \end{split}
\end{equation}
The term $\theta_{XYZ}^{(1)}(t)$ contains the factor $\E^{\I l \omega_0 t}$ which describes energy exchange with phonons. As polaron band transport is coherent, phonon scattering processes are absent. Consequently, only the term $\theta_{XYZ}^{(0)}$ is relevant for polaron band transport and we include only this term in what follows. Next, we use the Kubo formula [Eq.~\eqref{kubo_formula}], we exploit the relations:
\begin{equation}
 \expval{a_{k_1}^\dagger\qty(t)a_{k_1}}
 =\int \dd\omega \E^{\I\omega t}
 {\tilde{A}_{k_1}\qty(\omega)}\E^{-\beta\omega}
\end{equation}
\begin{equation}
 \expval{a_{k_2}\qty(t)a_{k_2}^\dagger}
 =\int \dd\omega e^{-\I \omega t}
 {\tilde{A}_{k_2}\qty(\omega)}
\end{equation}
where $\tilde{A}$ is the polaron spectral function for the Hamiltonian $\tilde{H}$ and we make use of the fact that $I_0\left[ 2a_{0,Z,X,Z+Y}\sqrt{n_{ph}(n_{ph}+1)}\right]\approx 1$ in the polaron band transport regime which takes place at low temperature when $n_{ph}$ is close to zero. We then arrive at the expression
\begin{equation}
    \mu_{\mathrm{dc}}={\pi \beta}\frac{\sum\limits_k \tilde{v}_k^2\int \mathrm{d}\omega \tilde{A}_k(\omega)^2 \E^{-\beta\omega}}{\sum\limits_k \tilde{n}_k} \phantom{.},
\end{equation}
with $\tilde{v}_k=-2\tilde{J}\sin k$ and $\tilde{n}_k=\E^{-\beta\tilde{\varepsilon}_k}$.
Subsequently, we can apply the approximation valid in situations where the spectral function is narrow, as it is here:
\begin{equation}
    \int \mathrm{d}\omega \tilde{A}_k(\omega)^2 \E^{-\beta\omega} =-\dfrac{\E^{-\beta\omega}}{2\pi\Im \Sigma_k(\omega)}\Big\vert_{\omega=\tilde{\varepsilon}_k}=\dfrac{\tilde{n}_k\tilde{\tau}_k}{\pi} \phantom{.}.
\end{equation}
with the polaron lifetime defined as $\tilde{\tau}_k=\frac{1}{-2 \Im \Sigma_k(\omega)}\Big\vert_{\omega=\tilde{\varepsilon}_k}$. The expression for self energy $\Sigma_k(\omega)$ is derived in Sec.~S2 in the Supplemental Material \cite{supplement}.
Finally, the dc mobility for polaron band transport can be expressed using the following formula:
\begin{equation}
    \label{mu_bt}
    \mu_{\mathrm{dc}}=\beta\dfrac{\sum_k \tilde{n}_k \tilde{\tau}_k \tilde{v}_k^2 }{\sum_k \tilde{n}_k}\phantom{.},
\end{equation}
which takes exactly the same form as the formula for conventional band transport [Eq.~\eqref{mu_wl0}] with a difference that the band energies, band velocities, state occupations and lifetimes should be replaced by their renormalized values. {We are not aware of any previous rigorous derivation of this formula.}

To evaluate the imaginary-time current-current correlation function in the polaron band transport regime, we make a further approximation that the spectral function can be approximated with the delta function
$\tilde{A}_k(\omega)\approx \delta (\omega-\tilde{\varepsilon}_k)$.
We note that such an approximation would not be good enough for real-time correlation function and the mobility since it does not lead to decay of correlation function at long real times but it is sufficient for imaginary times. The imaginary-time current-current correlation function is then given as
\begin{equation}\label{eq:C_jjz_it}
    C_{jj}(z)=\dfrac{\sum_{k_1 k_2}\E^{-\beta\tilde{\varepsilon}_{k_1}}\E^{\I(\tilde{\varepsilon}_{k_1}-\tilde{\varepsilon}_{k_2})z}Y_{k_1 k_2}^{(0)}}{\sum_k\E^{-\beta\tilde{\varepsilon}_k}}\phantom{.},
\end{equation}
where $z=-\I\tau$ ($0\le\tau\le \beta$) and $Y_{k_1 k_2}^{(0)}$ denotes the contribution to $Y_{k_1 k_2}$ from the term $\theta_{XYZ}^{(0)}$.

\subsubsection{Selected results in different transport regimes}\label{sec:2c4}

In this section, we present several results obtained using approximate formulas for the dc mobility and the correlation function presented in previous sections. With this, we get an initial insight about the transport regimes in different parts of parameter space. To present the results, we express all energies in terms of the electronic transfer integral $J$, while we express the electron-phonon interaction strength through the dimensionless parameter $\lambda=G^2/(2J\omega_0)$.

\begin{figure}[!h] 
  \begin{subfigure}[b]{0.95\linewidth}
    \centering
    \includegraphics[width=0.9\textwidth]{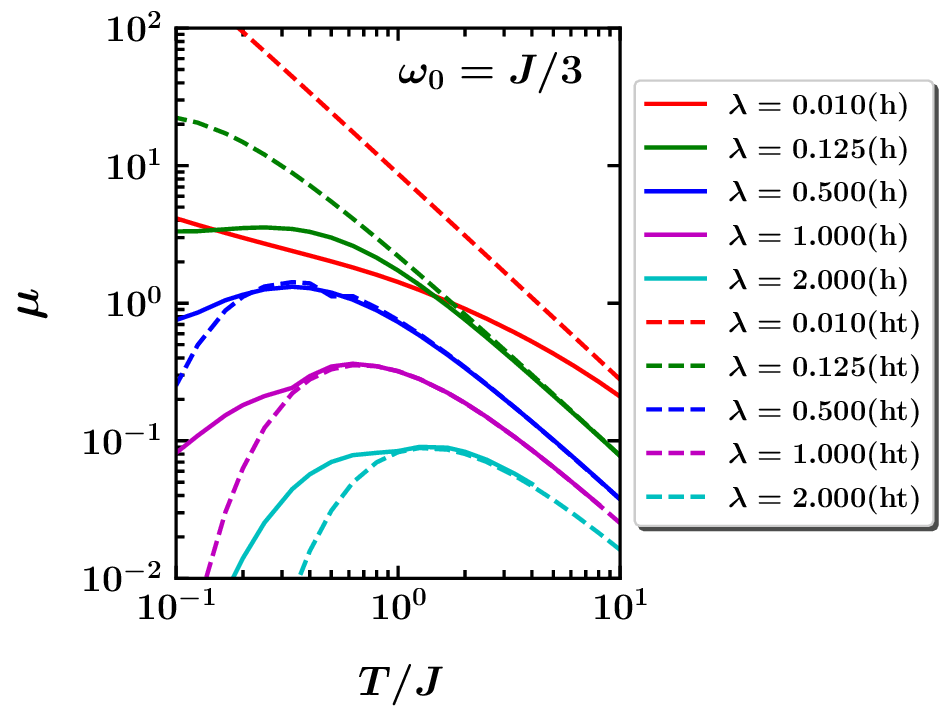}
    \caption{} 
    \label{slika_1a} 
    \vspace{4ex}
  \end{subfigure} 
  \begin{subfigure}[b]{0.95\linewidth}
    \centering
   \includegraphics[width=0.9\textwidth]{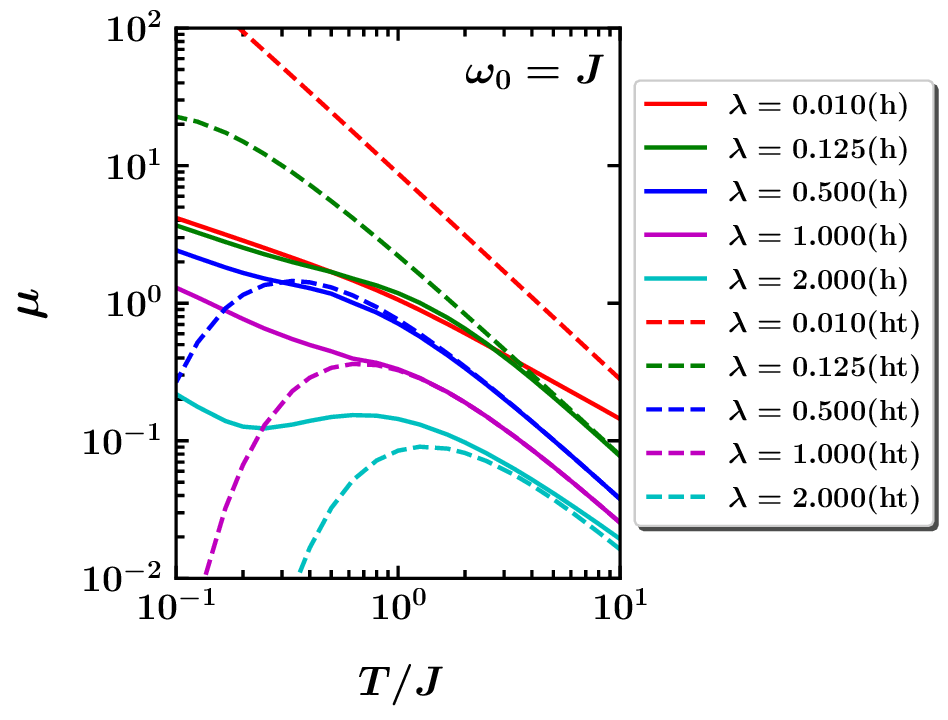}
    \caption{} 
    \label{slika_1b} 
  \end{subfigure}
  \caption{Hopping mobility obtained using Eq.~\eqref{mu_slA} is shown with full lines and the label "h", while the mobility obtained from Eq.~\eqref{mu_slhtA} in the high-temperature limit is shown with dashed lines and the label "ht". Results are shown for a range of temperatures and various interaction strengths $\lambda$. The results for phonon angular frequency $\omega_0=J/3$ are shown in part (a), while the results for $\omega_0=J$ are shown in part (b) of the figure.}
  \label{fig1} 
\end{figure}

{We start with the hopping regime and discuss the connection between the two different formulas in that regime.}
{In Sec. \ref{sec:2c1} we have presented expressions for hopping mobility and current-current correlation function [Eqs. \eqref{mu_slA} and \eqref{Cjj_slA}] in the case of the Holstein model with a single dispersionless phonon mode with correction that gives finite dc mobility in whole temperature range. We are also considering expressions for dc mobility [Eq. \eqref{mu_slhtA}] and current-current correlation function [Eq. \eqref{Cjj_slhtA}] derived from Marcus formula for hopping between two sites which are valid in high-temperature limit. It is expected that the expressions derived within these two approaches give the same result at high enough temperatures.}

\begin{figure}[!h] 
  \begin{subfigure}[b]{0.95\linewidth}
    \centering
   \includegraphics[width=0.9\textwidth]{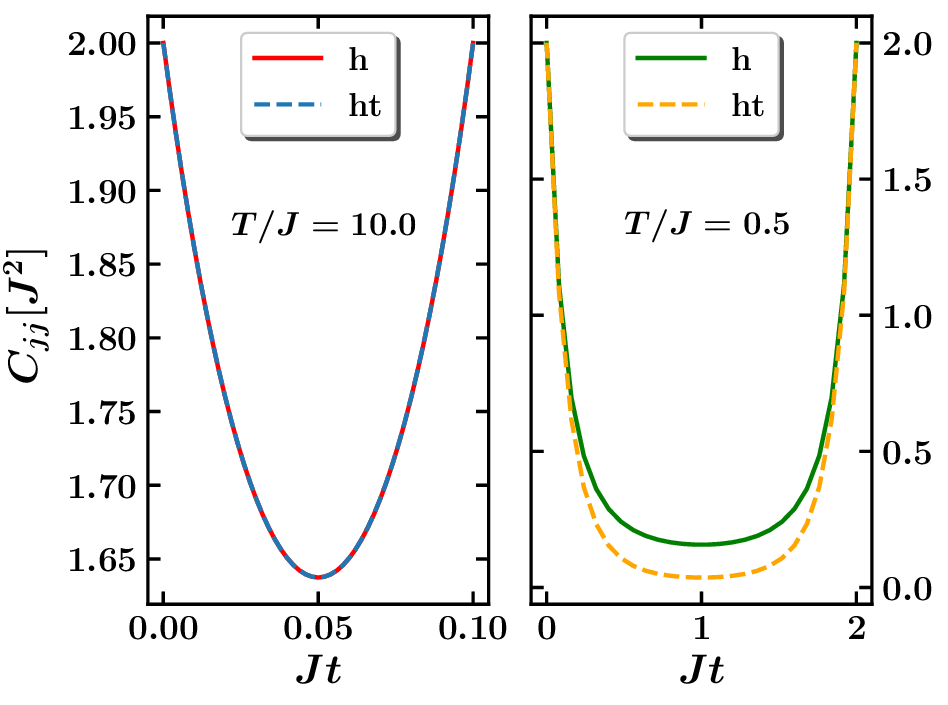}
    \caption{} 
    \label{slika_1d_im} 
    \vspace{4ex}
  \end{subfigure} 
  \begin{subfigure}[b]{0.95\linewidth}
    \centering
  \includegraphics[width=0.9\textwidth]{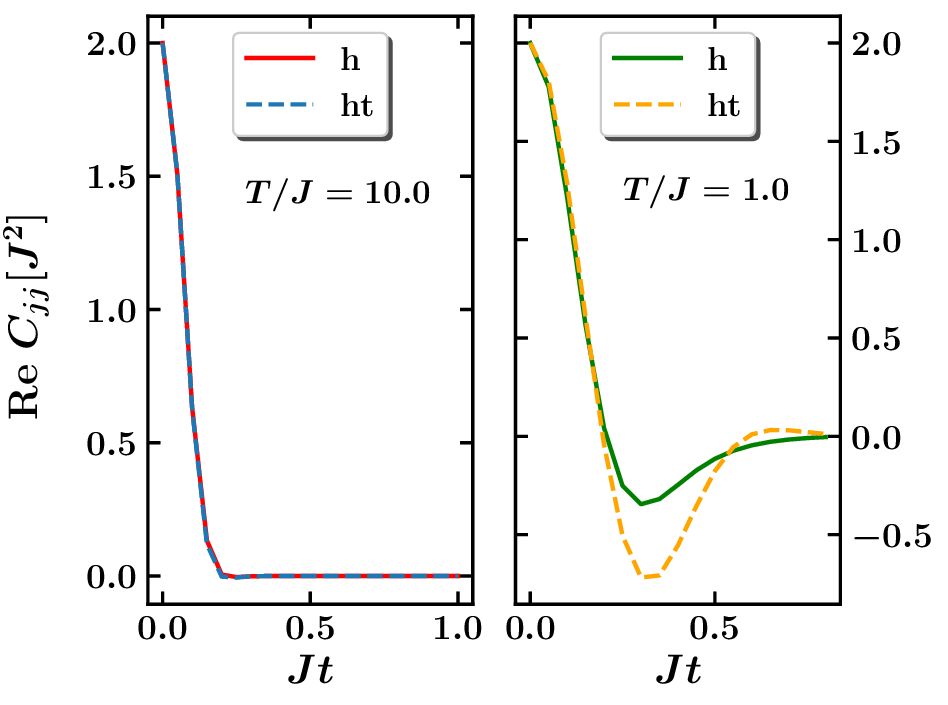}
    \caption{} 
    \label{slika_1d_re} 
  \end{subfigure}
  \caption{Current-current correlation function in imaginary time (a) and real part of the current-current correlation function in real time (b). Results shown with full lines labeled "h" are obtained with Eq. \eqref{Cjj_slA} and results obtained with Eq. \eqref{Cjj_slhtA} in the high-temperature limit are shown with dashed lines and labeled "ht". The results are shown for $\omega_0=3J$, $\lambda=2.000$ and the temperatures $T/J=10.0$ and $T/J=0.5$ for imaginary-time correlation functions and $T/J=10.0$ and $T/J=1.0$ for real time correlation functions.}
  \label{fig2} 
\end{figure}

From Fig. \ref{fig1} we can see that the curves for mobility obtained with Eq.~\eqref{mu_slA} and Eq.~\eqref{mu_slhtA} tend to get closer with increasing temperature and eventually converge towards each other when the temperature is high enough.
From Fig. \ref{fig2} we can see that current-current correlation functions follow the same trend as mobility. Mainly, we can see that correlation functions in imaginary time, as well as real parts of correlation functions in real time, match at high temperature and do not match at low temperature. It is also worth commenting on the effect of the phonon frequency. From Fig. \ref{slika_1a} and Fig. \ref{slika_1b} it is obvious that as the phonon energy $\omega_0$ is larger compared to the transfer integral $J$, the high-temperature limit is reached at higher temperatures. 
{The same correspondence can be made regarding the electron-phonon interaction strength; the stronger the interaction $\lambda$ the high-temperature effects take place at higher temperatures. We must pay attention to the fact that all expressions discussed in this paragraph are derived under the assumption of hopping transport and have physical sense only for stronger electron-phonon interactions ($\lambda \geq 0.500$).}
This analysis gives us confidence in the formulas {for dc mobility given in Eq. \eqref{mu_slA} and curent-current correlation function given in Eq. \eqref{Cjj_slA}} derived in Sec. \ref{sec:2c1} with the assumption of strong electron-phonon interaction and small electron transfer integral since they coincide with the {well-known} Marcus formula in the high-temperature limit.

\begin{figure}[htbp]
  \begin{subfigure}[b]{0.95\linewidth}
    \centering
   \includegraphics[width=0.9\textwidth]{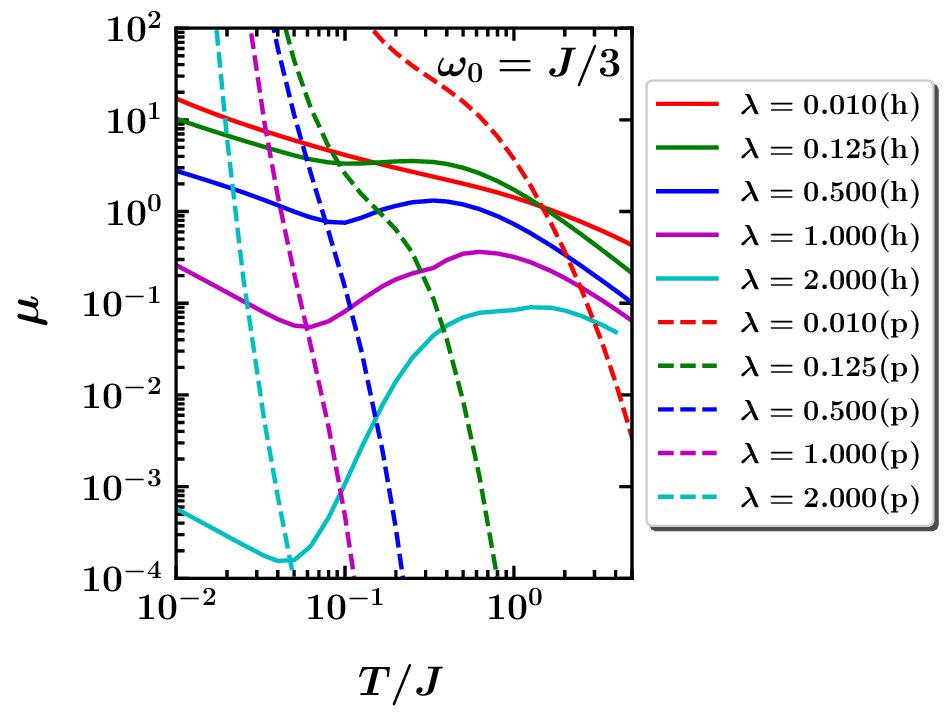}
    \caption{} 
    \label{slika_2a} 
    \vspace{4ex}
  \end{subfigure} 
  \begin{subfigure}[b]{0.95\linewidth}
    \centering
   \includegraphics[width=0.9\textwidth]{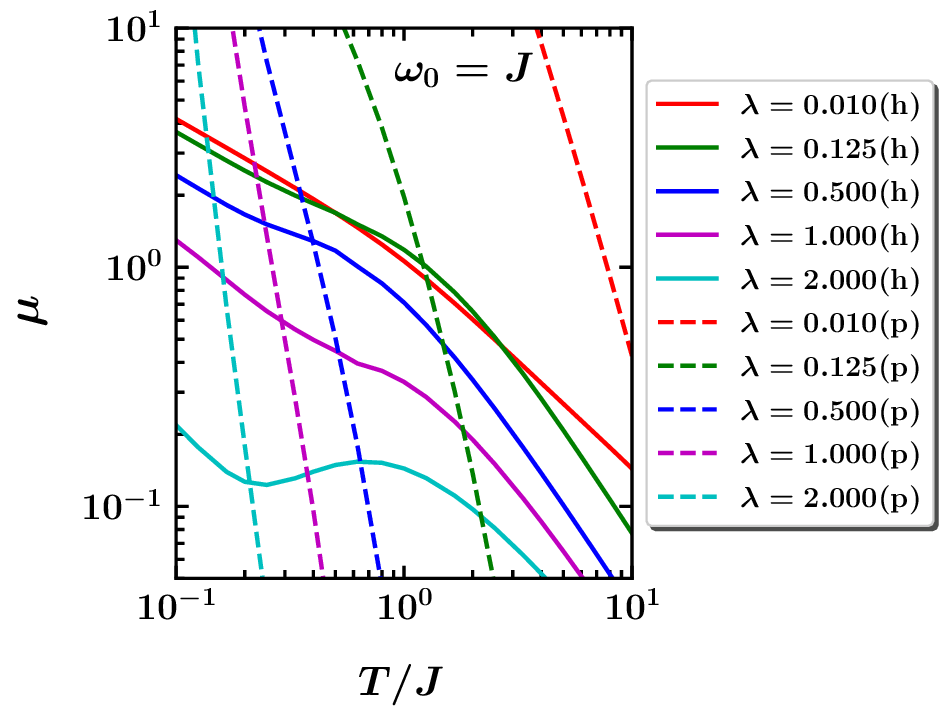}
    \caption{} 
    \label{slika_2b}
    \vspace{4ex}
  \end{subfigure}
  \begin{subfigure}[b]{0.95\linewidth}
    \centering
   \includegraphics[width=0.9\textwidth]{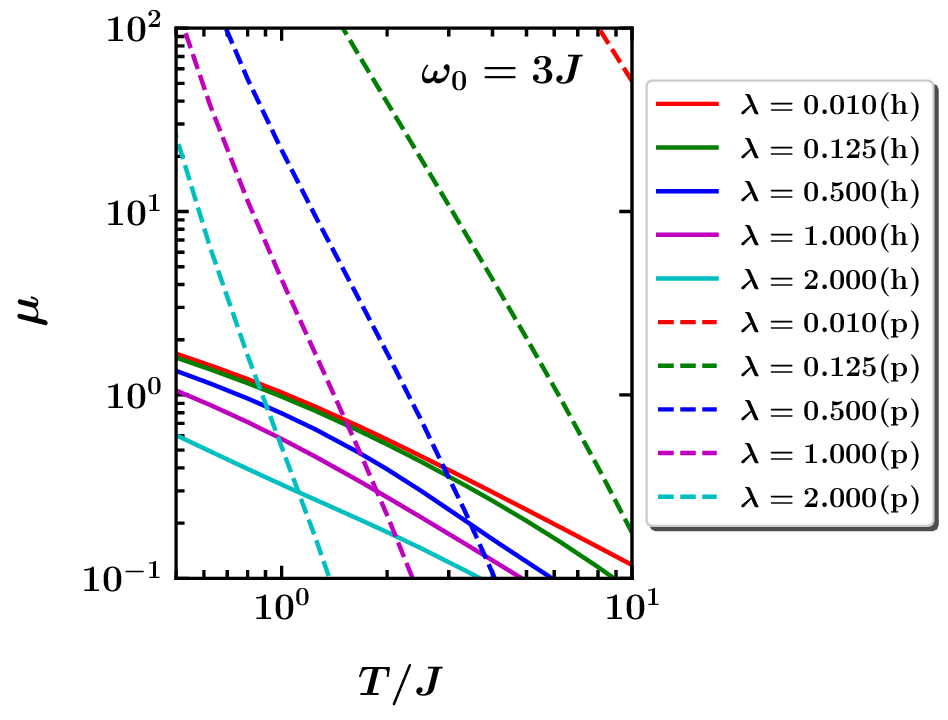}
    \caption{} 
    \label{slika_2c} 
  \end{subfigure}
  \caption{Mobility obtained from Eq. \eqref{mu_slA} for hopping is shown with full line and labeled "h". Mobility for polaron band transport is shown with dashed line and labeled "p". Results are shown for various interaction strengths $\lambda$ and for phonon angular frequencies $\omega_0=J/3$ in part (a), $\omega_0=J$ in part (b)  and $\omega_0=3J$ in part (c).}
  \label{fig3} 
\end{figure}

{Next, we discuss the polaron band transport regime and its crossover to the hopping regime.}
In Sec. \ref{sec:2c2} we discussed that at very low temperatures and strong electron-phonon interactions it is expected that polaron band transport will take place. To see exactly at how low temperatures this type of transport occurs, we will compare the results obtained using Eq. \eqref{mu_bt} for polaron band transport with the results obtained from Eq. \eqref{mu_slA} for hopping. The intersection of these curves can be considered a good estimate at which temperature the transition from polaron band transport to hopping transport occurs.
As we can see in Fig. \ref{fig3}, the greater the interaction strength, the intersection of the polaron band transport and the hopping transport mobilities is at lower temperature. We can conclude that, for stronger electron-phonon interactions, the hopping transport will occur at lower temperatures. Also, as we already saw in the analysis before this, with increasing phonon frequency the high-temperature effects arise at higher temperatures. This can be seen in Fig. \ref{fig3} as for $\omega_0=3J$ the intersections are at higher temperatures compared to the case when $\omega_0=J/3$. While the results for all interaction strengths are presented in Fig.~\ref{fig3}, one should bear in mind that only the results for stronger interactions (say $\lambda\ge 0.5$) should be considered for the analysis of the crossover from polaron band transport to hopping. {This analysis gave us an important insight into the range of parameters where we can expect the polaron band transport regime to hold.}

\begin{figure}[htbp]
    \centering
   \includegraphics[width=0.45\textwidth]{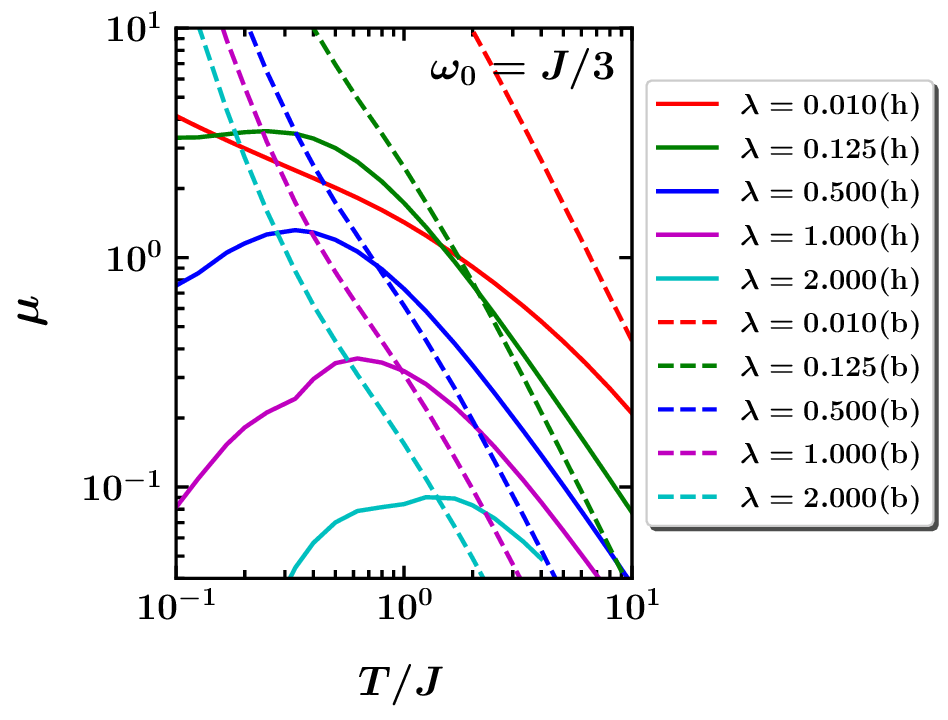}
  \caption{Temperature dependence of mobility shown for a range of interaction strengths. Full line labeled as "b" stands for band transport and dashed line labeled as "h" denotes hopping mobility. Results are shown for a range of interaction strengths $\lambda$ and for phonon angular frequency $\omega_0=J/3$.}
  \label{fig4} 
\end{figure}

{Finally, we discuss the band transport regime and its crossover to the hopping regime}.
{In Sec. \ref{sec:2c2} we presented expressions for the current-current correlation function [Eq. \eqref{Cjj_wl_1a}] and the dc mobility [Eq. \eqref{mu_wl0}] in the band-transport regime. These expressions are derived under the assumption of weak electron-phonon interaction. The carrier in this transport regime is a large polaron as opposed to small polaron that is characteristic for hopping transport. A large polaron moves similarly to a free carrier but with increased effective mass and it sometimes  scatters on thermaly induced lattice phonons. At high temperatures due to increased lattice vibrations the large polaron may turn into a small polaron.}
{Therefore, we} may also expect that even in the case of weak electron-phonon interaction, the crossover from band transport to hopping could occur at high enough temperatures. Hopping mobility at higher temperatures experiences a slower decline compared to a fast decrease in band-type mobility as a result of increased phonon scattering with temperature.
As can be seen in Fig. \ref{fig4}, with an increase in temperature (for a relatively weak electron phonon interaction), the hopping mobility tends toward band mobility and they eventually intersect.  Although the results for all interaction strengths are presented in Fig.~\ref{fig4}, one should bear in mind that only the results for weaker interaction (say $\lambda \leq 0.500$) should be considered for the analysis of the crossover from band transport to hopping.

Previous analysis gives us overall expectation for the validity range of relevant transport regimes throughout the parameter space. Namely, at extremely low temperatures and strong electron-phonon interaction, polaron band transport regime is prevailing. In contrast, when the temperature is high and there is a strong electron-phonon interaction, hopping transport is the dominant mechanism. For weak electron-phonon interactions, conventional band transport is generally the dominant transport mechanism except possibly at very high temperatures. To put these expectations on a solid ground, it is essential to compare our findings, derived from equations presented in Sec. \ref{sec:2c}, with numerically exact results obtained without approximations and assumptions of the transport regime.

\section{Results}\label{sec:3}

To reliably identify the transport regimes throughout the parameter space of the Holstein model, we compare numerically exact results for the imaginary-time current-current correlation function obtained using path-integral Quantum Monte Carlo with approximate results obtained under the assumption of each of the transport regimes. We performed QMC calculations throughout the parameter space $0.1 \le T/J \le 10.0$, $0.01 \le \lambda \le 2.0$ for three values of the phonon angular frequency $\omega_0=J/3$, $\omega_0=J$ and $\omega_0=3J$. While it would be very desirable to have the comparison at the level of the real-time correlation functions or the dc mobility, it is currently not possible to obtain numerically exact results for such a broad range of model parameters. For a limited set of model parameters where numerically exact results for the dc mobility are available, we make such a comparison. Namely, we compare the dc mobility for the three transport regimes with the numerically exact results obtained using the hierarchical equations of motion (HEOM) method in Ref.~\cite{jcp159-094113} and with our previous results obtained from real- and imaginary-time QMC in Ref.~\cite{prb107-184315}.


\subsection{Comparison of imaginary-time current-current correlation functions}

For the comparison of imaginary-time current-current correlation function with the {correlation functions} in the band transport and in the hopping regime we define a numerical criterion that describes how much imaginary-time correlation functions differ from each other. The numerical value that describes how much the function $C_{jj}^{\mathrm{(a)}}(t)$ deviates from the numerically exact function $C_{jj}^{\mathrm{QMC}}(t)$ is given by the formula:
\begin{equation}
    \mathcal{D}_{C_{jj}}=\dfrac{\int \mathrm{d}t \phantom{.}\Big\vert C_{jj}^{\mathrm{QMC}}(t)-C_{jj}^{\mathrm{(a)}}(t)\Big\vert}{\Big\vert \int \mathrm{d}t\phantom{.}C_{jj}^{\mathrm{QMC}}(t)\Big\vert}\phantom{.},
\end{equation}
where we have taken $C_{jj}^{\mathrm{QMC}}$ to be our reference value. We consider that the two functions {are in good agreement} if $\mathcal{D}_{C_{jj}}<0.2$.

In the case of the polaron band transport regime, we choose a different criterion based on the following physical arguments. In the polaron band transport regime, all physically relevant processes happen in a narrow range of energies whose width is given by the renormalized bandwidth $4\tilde{J}$. In the derivations of polaron band transport equations in Sec.~\ref{sec:2c2} this range of energies was effectively selected by choosing only the $\theta_{XYZ}^{(0)}$ term in relevant equations. On the other hand, exact current-current correlation function contains also the information about the processes of multiphonon emission/absorption that happen at energies equal to multiples of the phonon energy. Therefore, one cannot expect an agreement between the {numerically} exact current-current correlation function $C_{jj}^{\mathrm{QMC}}$ and the  one obtained under the assumption of polaron band transport $C_{jj}^{(\mathrm{p})}$, even when the system is in the polaron band transport regime. On the other hand, as discussed in Refs. \cite{prb76-035115,prb82-165125}, the $C_{jj}$ at imaginary-time $t=-\I\frac{\beta}{2}$ contains only the information about the conductivity at low frequencies, which is exactly the frequency range relevant for polaron band transport. Consequently, to identify if the system is in the polaron band transport regime, we evaluate the quantity
\begin{equation}\label{eq:dcjjpb}
    \mathcal{D}_{C_{jj}}=\dfrac{\Big\vert C_{jj}^{\mathrm{QMC}}(t=-\I\frac{\beta}{2})-C_{jj}^{(\mathrm{p})}(t=-\I\frac{\beta}{2})\Big\vert}{\big\vert C_{jj}^{\mathrm{QMC}}(t=-\I\frac{\beta}{2})\big\vert}\phantom{,}.
\end{equation}
When the system is in the polaron band transport regime, the $\theta_{XYZ}^{(0)}$ term dominates over the $\theta_{XYZ}^{(1)}$ term in equations in Sec.~\ref{sec:2c2} and therefore $\mathcal{D}_{C_{jj}}$ from Eq.~\eqref{eq:dcjjpb} satisfies $\mathcal{D}_{C_{jj}}<0.5$. Hence, we consider that the system is in the polaron band transport regime if $\mathcal{D}_{C_{jj}}<0.5$, with $\mathcal{D}_{C_{jj}}$ given in Eq.~\eqref{eq:dcjjpb}.

We further note that the formulas for imaginary-time $C_{jj}$ in the polaron band transport and in the conventional band transport regime converge to the same value when one sets the electron-phonon interaction to be small. Namely, for small interaction, $Y_{k_1 k_2}$ in Eq.~\eqref{eq:yk1k2} reduces to $Y_{k_1 k_2}=4J^2\delta_{k_1k_2}\sin^2 k_1$ (because in this limit $\tilde{J}=J$ and $\theta_{XYZ}=1$) and $C_{jj}(z)$ from Eq.~\eqref{eq:C_jjz_it} then reduces to
\begin{equation}\label{eq:C_jjz_it2}
    C_{jj}(z)=4J^2\dfrac{\sum_{k}\E^{-\beta{\varepsilon}_{k}}\sin^2k}{\sum_k\E^{-\beta{\varepsilon}_k}}\phantom{.}.
\end{equation}
The same result is obtained from Eq.~\eqref{Cjj_wl_1a} by exploiting that for small interaction
$\langle a_k^{\dagger}(t)a_k \rangle = \E^{-\qty(\beta-\I t)\varepsilon_k}$,
$\langle a_k(t)a_k^{\dagger} \rangle = \E^{-\I\varepsilon_k t}$. Hence, the polaron band transport formula contains the band transport formula as its special case. For this reason, whenever our criterion for $\mathcal{D}_{C_{jj}}$ suggests that both polaron band transport and conventional band transport are possible, we assign the conventional band transport regime for these values of parameters.

\begin{figure}[htbp]
  \begin{subfigure}[b]{0.95\linewidth}
    \centering
    \includegraphics[width=0.9\textwidth]{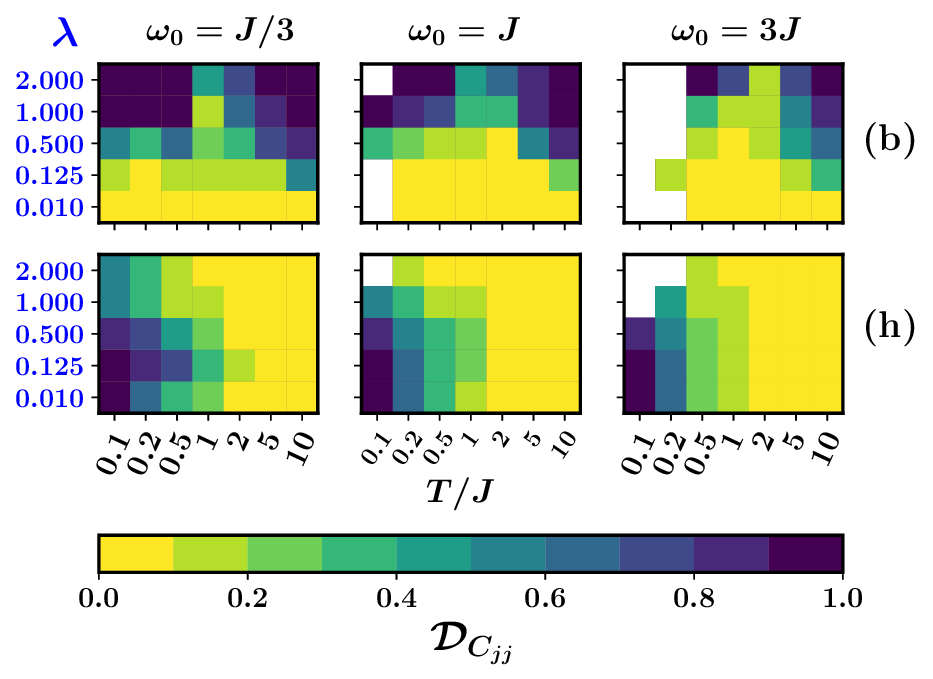}
    \caption{} 
    \label{cmap_a} 
    \vspace{4ex}
  \end{subfigure} 
  \begin{subfigure}[b]{0.95\linewidth}
    \centering
   \includegraphics[width=0.9\textwidth]{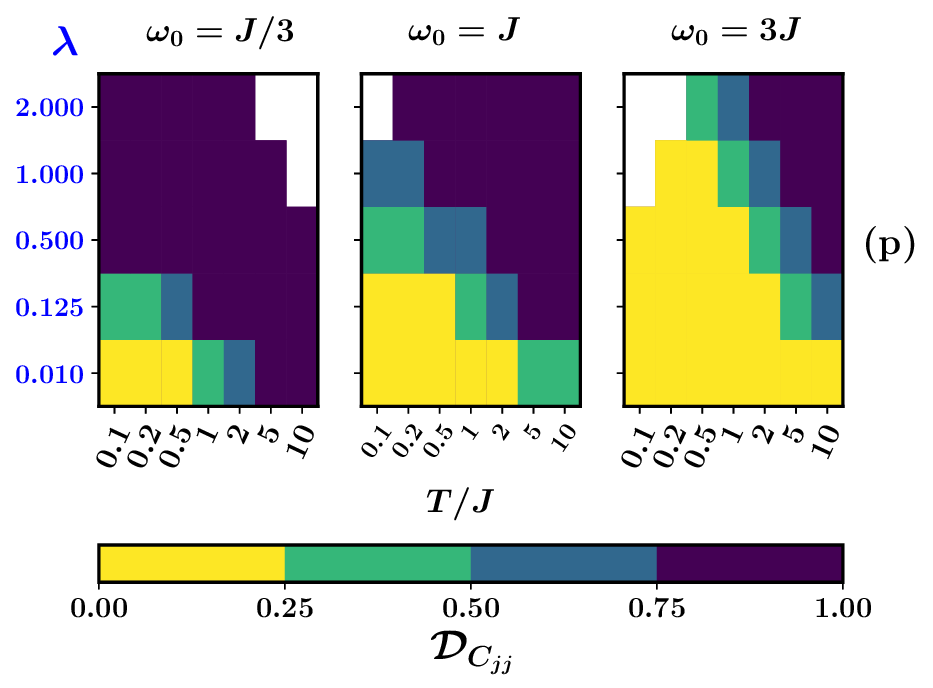}
    \caption{} 
    \label{cmap_b} 
  \end{subfigure}
  \caption{The relative difference $\mathcal{D}_{C_{jj}}$ of the approximate and QMC imaginary-time current-current correlation functions for different values of the temperature $T$, the interaction strength $\lambda$ and the phonon angular frequency $\omega_0$. The results are presented for three phonon frequency values across the three respective columns. Each row represents the difference for the specified transport regime. In part (a), the results for band transport (b) and hopping transport (h) are presented, while in part (b), the result for polaron band transport (p) is presented.}
  \label{cmap} 
\end{figure}

\begin{figure}[!h] 
    \centering
   \includegraphics[width=0.45\textwidth]{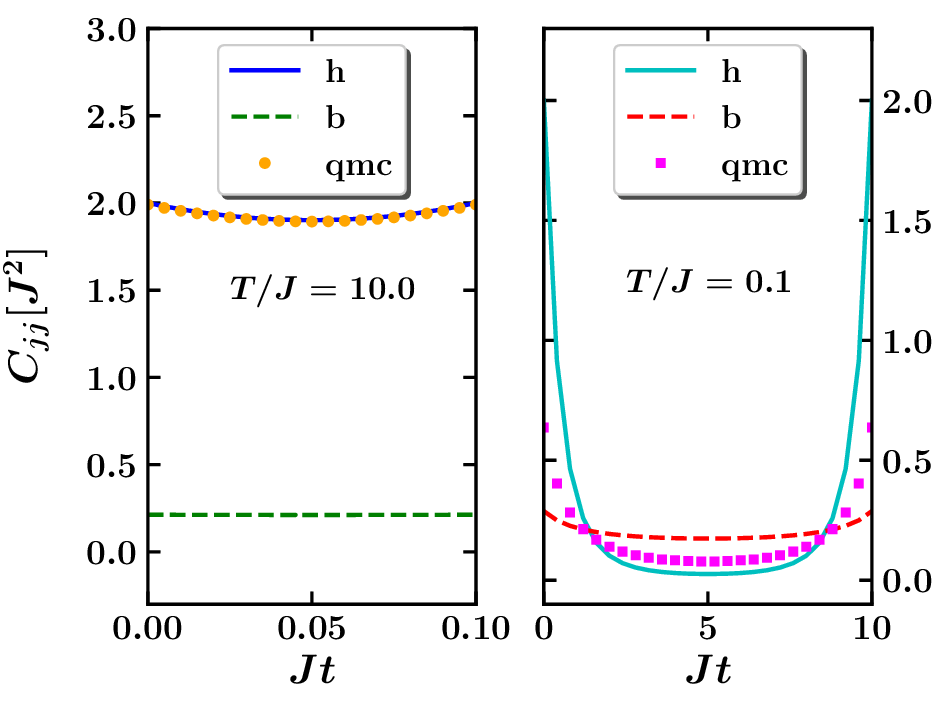}
  \caption{ Current-current correlation functions in imaginary time at temperatures $T/J=10.0$ (a) and $T/J=0.1$ (b) for the parameters $\omega_0=J/3$ and $\lambda=0.500$. Correlation functions obtained using Eq.~\eqref{Cjj_slA} for hopping transport are presented in full lines and labeled "h", while the functions calculated using formulas from Sec. \ref{sec:2c3} for band transport are presented in dashed lines and labeled "b". QMC results are represented with points.
  }
  \label{slika_5} 
\end{figure}

{The results for the quantity $\mathcal{D}_{C_{jj}}$ for all the three regimes are presented in Fig. \ref{cmap}.} As previously stated, for both hopping and band transport, we have set the threshold for tolerance between functions at a maximum of $20\%$. To demonstrate what this means, we  show an example of current-current correlation functions for parameter values ($\omega_0=J/3$, $\lambda=0.500$) in Fig. \ref{slika_5}. As can be seen in Fig. \ref{slika_5}, there is an obvious matching of the QMC data with the function obtained for hopping transport at a temperature $T/J=10.0$ which is also visible in Fig. \ref{cmap} where numerical value $\mathcal{D}_{C_{jj}}$ is less than $0.2$ in case of hopping transport. On the other hand, neither band transport nor hopping results match the QMC data at a temperature $T/J=0.1$, as can be seen from Fig. \ref{slika_5}.
This is also evident in Fig. \ref{cmap_a}, where the numerical value $\mathcal{D}_{C_{jj}}$ is greater than $0.2$ for both band and hopping transport. This value is lower in the case of band transport; nevertheless, it remains approximately $0.5$, which is considerably higher than our set threshold of $0.2$.

Several points are missing in Fig.~\ref{cmap} for the following reasons. It is very challenging to perform QMC calculations at high interaction strengths and low temperatures, as discussed in Sec.~\ref{sec:2b}. Therefore, we could not obtain the data at the lowest temperatures and the strongest interactions for phonon frequencies $\omega_0=J$ and $\omega_0=3J$. {It is well known in the literature that it is challenging to obtain current-current correlation functions at low temperatures and strong interactions, as evidenced, for example, by the fact that numerically exact dc mobilities can be obtained in practice only at temperatures $T\gtrsim 1$ and interactions $\lambda\lesssim 1$ \cite{prb107-184315,jcp159-094113}.}
In the case of band transport calculations, the spectral function becomes very narrow at lowest temperatures which poses challenge for the calculation. This prevented us from obtaining converged results for band transport at the lowest temperatures for phonon frequencies $\omega_0=J$ and $\omega_0=3J$. {Similar issue was observed in calculations for weak coupling in Ref.~\cite{prb99-104304}}. Lastly, numerical issues in the calculation of Bessel functions arise in calculations that assume the polaron band transport regime for highest temperatures and strongest interactions when $\omega_0=J/3$ (where it is actually expected that this regime does not take place).

Fig. \ref{cmap} can be to a significant extent used to identify the relevant transport regimes throughout the parameter space. However, for certain parameters, we see from the figure that more than one regime is possible in principle. In such cases, to distinguish which regime is in place, we complement our analysis with {dc mobilities and real-time current-current correlation functions} when these results are available.

\subsection{Comparison of mobilities}

{In this section we compare dc mobilities obtained using the expressions for hopping given by Eq. \eqref{mu_slA}, polaron-band transport given by Eq. \eqref{mu_bt} and conventional band transport given by Eq. \eqref{mu_wl0} with numerically exact mobilities that are obtained using HEOM and path-integral QMC techniques.}

\begin{figure}[htbp]
  \begin{subfigure}[b]{0.95\linewidth}
    \centering
    \includegraphics[width=0.9\textwidth]{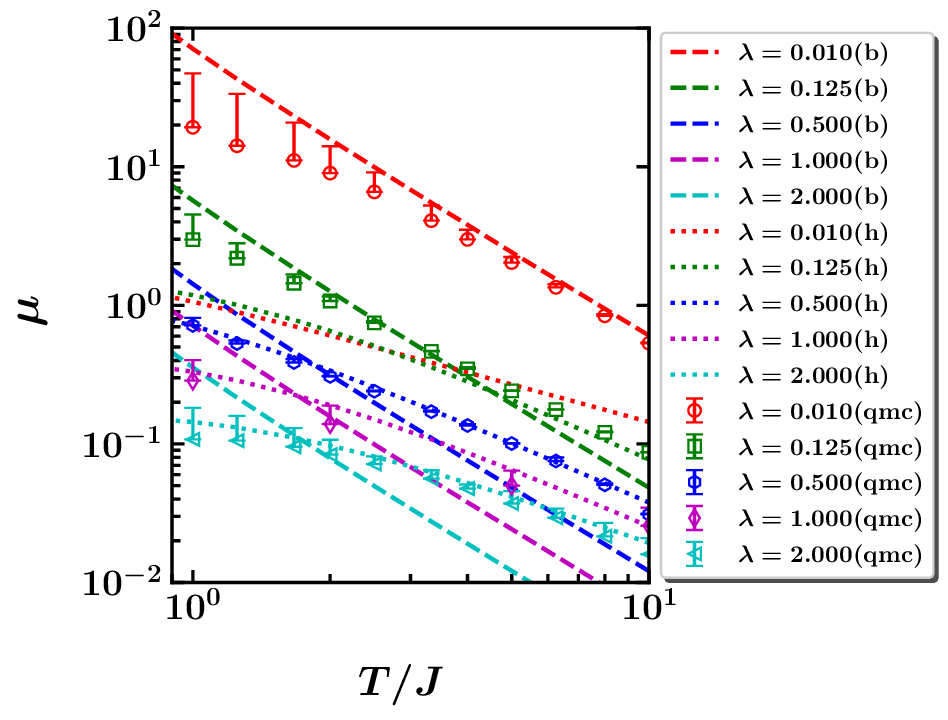}
    \caption{} 
    \label{mobility_1.0_qmc} 
    \vspace{4ex}
  \end{subfigure} 
  \begin{subfigure}[b]{0.95\linewidth}
    \centering
   \includegraphics[width=0.9\textwidth]{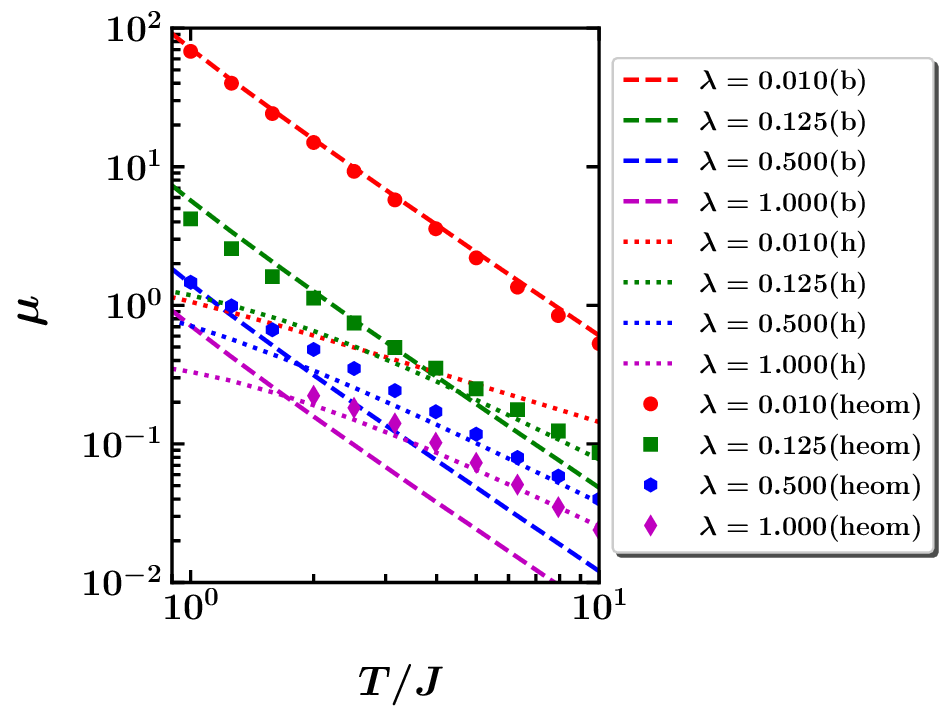}
    \caption{} 
    \label{mobility_1.0_heom} 
  \end{subfigure}
  \caption{Temperature dependence of mobility obtained using formulas for hopping (labeled as "h") and band transport (labeled as "b") compared with numerically exact QMC data from Ref.~\cite{prb107-184315} (a) and HEOM data from Ref.~\cite{jcp159-094113} (b). Numerical data are presented as points. The upward error bars in QMC data were estimated as described in Ref.~\cite{prb107-184315}.  The results are shown for $\omega_0=J$.}
  \label{mobility_1.0} 
\end{figure}

\begin{figure}[!h] 
    \centering
   \includegraphics[width=0.45\textwidth]{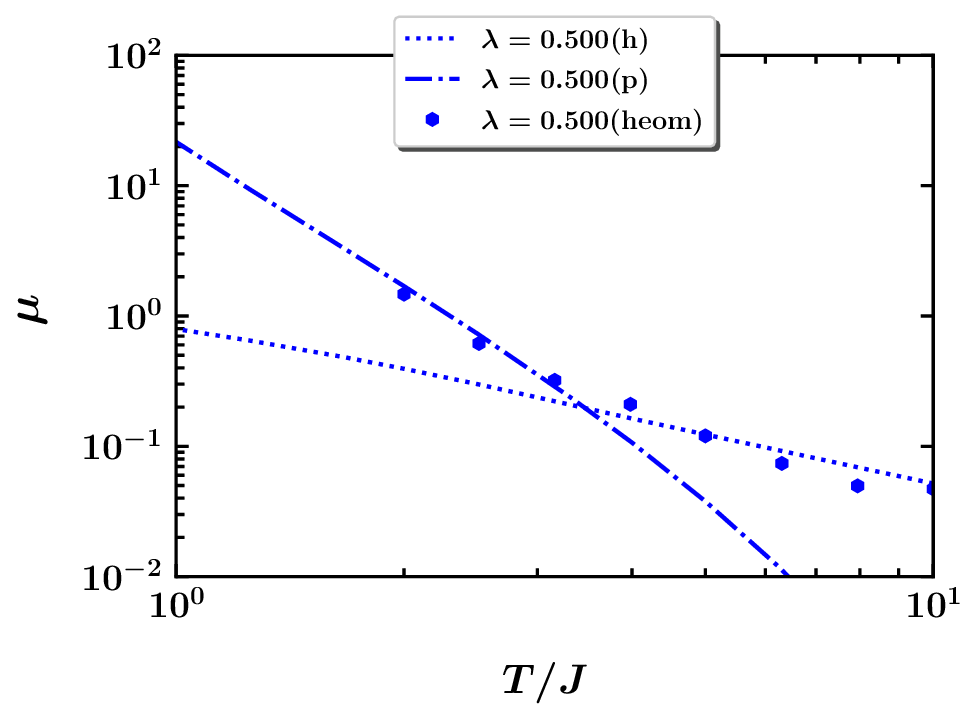}
   \caption{Temperature dependence of mobility obtained using formulas for hopping (labeled as "h") and polaron band transport (labeled as "p") compared with exact mobility obtained using the HEOM method in Ref.~\cite{jcp159-094113} (labeled as "heom"). The results are shown for $\omega_0=3J$ and $\lambda=0.500$.}
  \label{mobility_3.0} 
\end{figure}

Fig. \ref{mobility_1.0} shows the mobility calculated for different temperatures and coupling strengths for a phonon energy $\omega_0=J$. This value {of phonon frequency} was chosen because of the availability of the most comprehensive and reliable HEOM (Ref.~\cite{jcp159-094113}) and QMC results (Ref.~\cite{prb107-184315}) {for this chosen set of parameters}. The lines show {results obtained with expressions from Sec. \ref{sec:2c}}, whereas the points depict the numerical results obtained using HEOM and QMC techniques. As depicted in Fig. \ref{mobility_1.0}, there is a noticeable agreement between the numerical data and the results {obtained with formulas} for certain transport regimes. Specifically, our band transport expression defined by Eq.~\eqref{mu_wl0} holds true for the minimal interaction {($\lambda = 0.010$)} in Fig. \ref{mobility_1.0} and remains reasonably accurate for the next stronger interaction {($\lambda = 0.125$)}. Furthermore, in case of stronger interactions {($\lambda \geq 0.500$)}, the numerical data align well with the hopping mobility given in Eq.~\eqref{mu_slA}. The result in Fig. \ref{mobility_1.0} clearly indicates that for the weakest interaction and high temperatures, {the band transport is the correct regime. This is clear from the numerical data in Fig.~\ref{mobility_1.0}, which are in perfect alignment with the band mobility, and within the temperature range displayed, the band and hopping mobilities do not intersect. Meanwhile, we cannot make such a clear distinction between hopping and band transport regimes based on results in Fig. \ref{cmap_a}.}  The challenges associated with numerically exact calculations at extremely low temperatures prevent us from depicting the crossover from polaron band transport to hopping transport in Fig. \ref{mobility_1.0}. However, this crossover can be seen in Fig. \ref{mobility_3.0}, for the phonon angular frequency of $\omega_0=3J$. It is possible to observe this crossover for this value of $\omega_0$, since it takes place at higher temperature, as could have been expected based on the results of Fig.~\ref{fig3}(c). {This result aligns with the result shown in Fig. \ref{cmap_b} where it indicates the dominance of polaron band transport regime up to temperature $T/J=2.0$ and based on Fig. \ref{mobility_3.0} the crossover takes place betweent temperatures $T/J=2.0$ and $T/J=3.0$.}

\subsection{Transport regime diagram for the one-dimensional Holstein model}

\begin{figure}[htbp]
    \centering
    \includegraphics[width=0.45\textwidth]{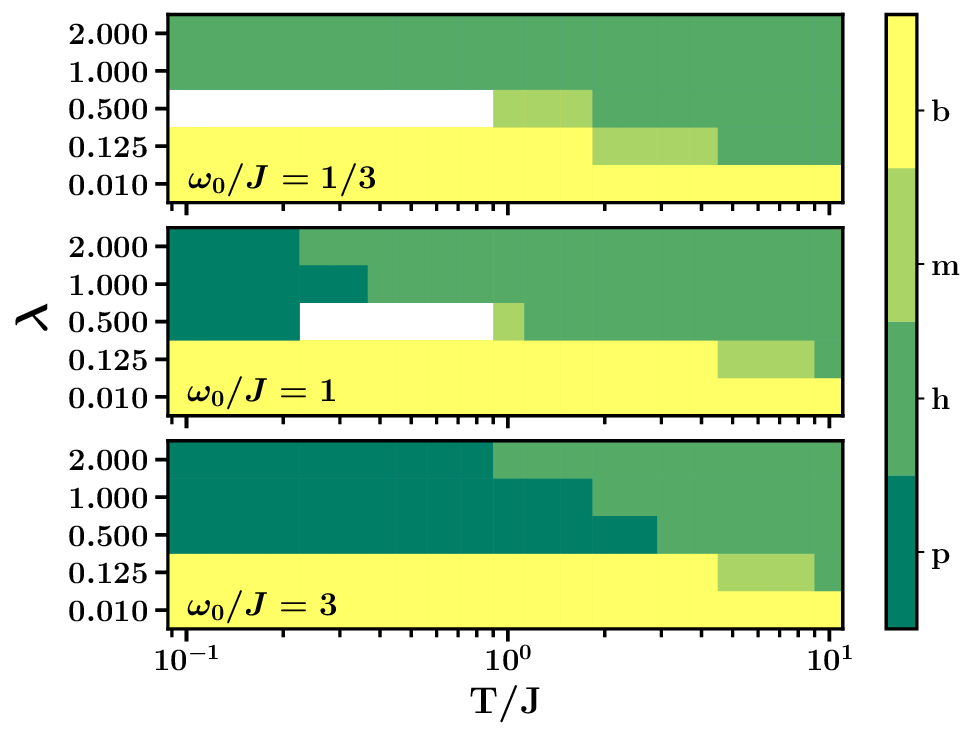}
  \caption{Transport regime diagram  for the Holstein model. The results are shown for three values of phonon angular frequency $\omega_0/J$ for the same values of the interaction strength $\lambda$ and the temperature $T/J$. Three distinct transport regimes are labeled as "b" for conventional band transport, "h" for hopping transport and "p" for polaron band transport. The label "m" stands for intermediate regime between band and hopping transport. {The unlabeled white areas correspond to areas in parameter space where none of the three transport regimes can be applied based on our results.}}
  \label{transport_regimes_corrected} 
\end{figure}

\begin{figure}[htbp]
    \centering
   \includegraphics[width=0.45\textwidth]{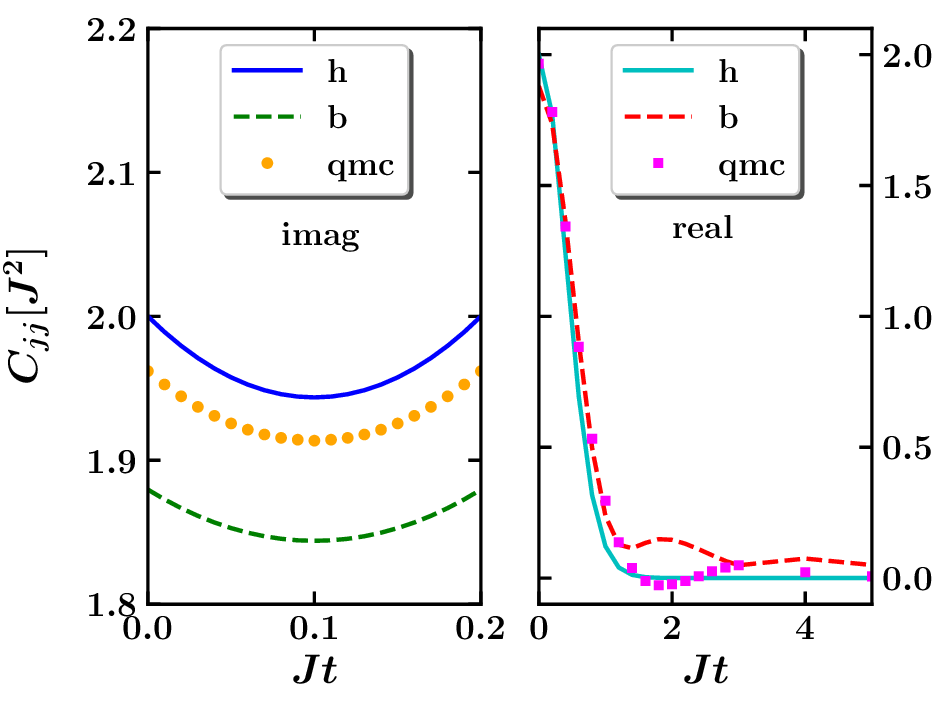}
  \caption{Imaginary- and real-time current-current correlation functions for $\omega_0=J$, $\lambda=0.125$, $T/J=5.000$. Lines labeled with "h" correspond to functions obtained with Eq.~\eqref{Cjj_slA} for hopping transport, lines labeled "b" correspond to functions obtained with expression for band transport and QMC results are represented with points.}
  \label{slika_5c}
\end{figure}

\begin{figure}[htbp]
    \centering
   \includegraphics[width=0.45\textwidth]{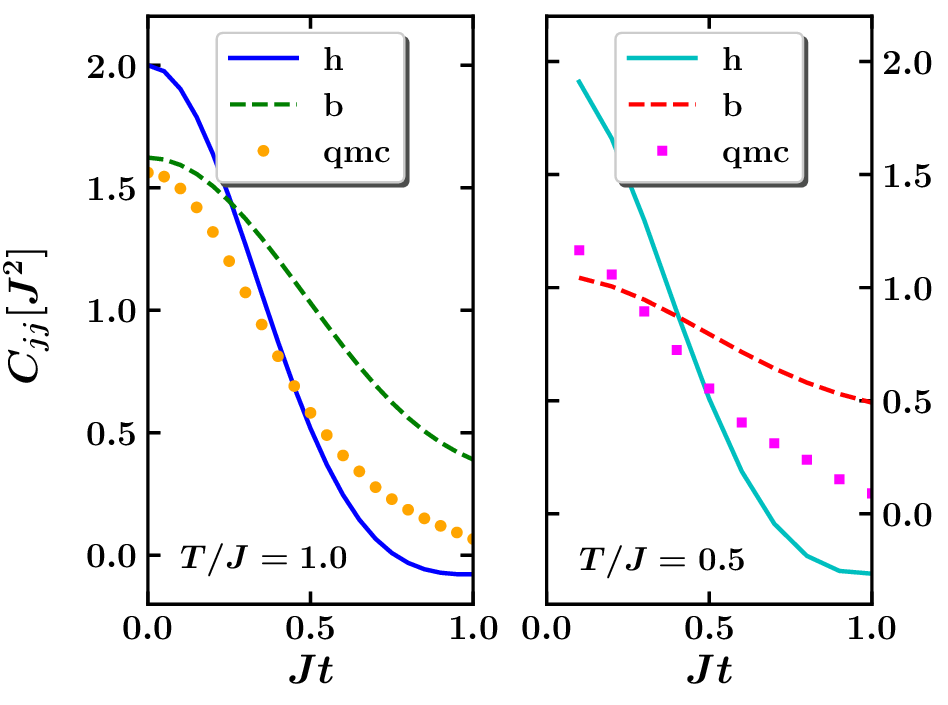}
  \caption{Real time current-current correlation functions for $\omega_0=J$, $\lambda=0.500$, and the temperatures $T/J=1.000$ and $T/J=0.500$. Lines labeled with "h" correspond to functions obtained for hopping transport, lines labeled "b" correspond to functions obtained with expression for band transport and QMC results are represented with points.}
  \label{slika_5d} 
\end{figure}

The final transport regime diagram based on all the {previous results} is presented in Fig.~\ref{transport_regimes_corrected}. In what follows we explain in detail how this diagram was {constructed} in the case when the phonon energy $\omega_0$ equals $J$, while we provide similar explanation for other values of $\omega_0$ in Secs. S3 and S4 in the Supplemental Material \cite{supplement}. It is evident from Fig.~\ref{cmap} that the band transport regime prevails in the areas characterized by the two weakest interactions and low temperatures. With an increase in temperature, Fig.~\ref{cmap} indicates that both band and hopping transport mechanisms could be possible. To clarify which transport regime is dominant at high temperatures, we turn our attention to Fig. \ref{mobility_1.0}. The figure clearly shows that across the entire temperature range for  the weakest interaction, the band transport regime prevails since numerically exact HEOM results fully agree with band transport results, while the curves for band and hopping mobility do not intersect. For the second interaction strength, $\lambda=0.125$, it remains difficult to differentiate between band and hopping mobility, as evident in Fig. \ref{mobility_1.0} where both expressions for dc mobility give a result similar to the numerically exact result. Similarly, Fig.~\ref{slika_5c} shows that distinguishing the real-time correlation functions for the band and hopping regimes is equally challenging in this range of parameters. For these reasons we depict this range of parameters as intermediate range at the crossover between band and hopping transport. We assign eventually the result at the highest temperature to the hopping regime since the imaginary-time result for the band regime starts to deviate from the numerically exact result, as seen in Fig.~\ref{cmap}. For the interaction strength $\lambda=0.500$ by looking at Fig. \ref{cmap} it is evident that hopping transport dominates at temperatures above $T/J=2.0$. At lowest temperatures (say $T/J=0.2$ and lower) one sees from Fig.~\ref{cmap} that the system is in the polaron band transport regime. For the intermediate range of temperatures between these extremes, Fig.~\ref{cmap} suggests that the system could be in the band transport regime. Fig.~\ref{slika_5d} shows that in this range of temperatures, real-time correlation functions obtained under the assumption of band transport are in better agreement with QMC results than the ones obtained under the assumption of hopping. Nevertheless, {there} is a noticeable discrepancy of band transport correlation functions and QMC results {for temperature $T/J=0.5$}. For these reasons we do not assign a single regime for these parameters and denote that this is an intermediate range of parameters {where none of mentioned regimes apply}. In Fig. \ref{transport_regimes_corrected}, {this} ambiguity is represented by a white area. {On the other hand, for a temperature of $T/J=1.0$, we classify it as an intermediate band transport-hopping regime which is depicted by label "m" in Fig. \ref{transport_regimes_corrected}.} This classification is due to the QMC correlation function results being {more closely} aligned with both the hopping and band transport correlation functions.
When the interaction strength is set to $\lambda=1.000$ and the temperature is $T/J=0.5$ or higher, Fig. \ref{cmap} clearly indicates that hopping is the appropriate transport regime. The crossover point for mobility curves of hopping and polaron band transport (see Fig.~\ref{fig3}) lies between temperatures $T/J=0.2$ to $T/J=0.3$. Therefore, we conclude that for temperatures below $T/J=0.3$, the polaron band transport regime is in place. For an interaction strength of $\lambda=2.000$ and temperatures less than $T/J=0.2$, we perform the same assignation for the same reason. {Based on previous analysis that combines results obtained with formulas from Sec. \ref{sec:2c} and numerically exact HEOM and QMC data for imaginary and real-time current-current correlation functions and dc mobilities, we construct the transport regime diagram in parameter space defined with interaction strength $\lambda$ and temperature $T/J$ for phonon frequency $\omega_0=J$ depicted in Fig. \ref{transport_regimes_corrected}. In the same manner we construct the diagrams for phonon frequencies $\omega_0=J/3$ and $\omega_0=3J$ with details given in Secs. S3 and S4 in the Supplemental Material \cite{supplement}.}

{Next, we discuss other works in the literature where transport regime diagrams of models with electron-phonon interaction were investigated. In Ref.~\cite{ncomms12-4260}, the authors considered a mixed Holstein-Peierls model relevant for organic semiconductors. They examined possible regimes when the strength of electronic coupling and non-local electron-phonon coupling are changed. Due to different type of Hamiltonian, and the fact that temperature was not varied in Ref.~\cite{ncomms12-4260}, these results are not directly comparable to ours. Nevertheless, band transport regime (termed band-like in Ref.~\cite{ncomms12-4260}) and the hopping regime (termed phonon-assisted in Ref.~\cite{ncomms12-4260}) were clearly identified, respectively, for large and small electronic coupling. In Ref. \cite{cphys6-125}, the authors considered a two site Holstein model and the crossover between different charge transfer regimes as they evolve over time for different adiabaticity ratios. They identify the regimes they term polaronic, soft-gating and transient localization. Our study offers complementary insight in comparison to these two works. We study the effects of temperature over a wide range which were not addressed in Ref.~\cite{ncomms12-4260}. While we address directly the long-range charge transport quantified by the dc mobilities, Ref. \cite{cphys6-125} was focused on short-range charge transfer.}

\section{Conclusion}\label{sec:4}

In conclusion, we studied in detail the transport regimes of a benchmark model with the electron-phonon interaction - the one-dimensional Holstein model. We computed the imaginary-time current-current correlation functions using numerically exact path-integral based QMC method for a broad range of model parameters that practically covers the whole parameter space. These were compared with the corresponding functions under the assumptions of the band transport regime, the hopping regime and the polaron band transport regime. The comparisons were used to establish the range of validity of each of these transport regimes. The analysis was complemented with comparison of real-time current-current correlation functions and dc mobilities for parameters where these data are available.

In accordance with expectations, the results indicate that band transport occurs at low interaction strengths and low temperatures, hopping transport is present for strong interaction and high temperatures, while polaron band transport takes place for strong interaction and low temperatures. More importantly, the results indicate that practically the whole parameter space is covered by the three mentioned transport regimes, except for some parts of the space at intermediate electron-phonon coupling strengths. This conclusion is something that one may not have expected. Such a conclusion might have important consequences for modeling charge transport in real materials. It is practically impossible  to model charge transport in real materials for arbitrary electron-phonon coupling strength without the assumption of a particular transport regime. On the other hand, as mentioned in Sec.~\ref{sec:1}, it is  possible to perform simulations of mobility in real materials assuming the band transport or the hopping transport regime. If our conclusions regarding the applicability of one of the three transport regimes throughout most of the parameter space could be extended from the benchmark Holstein model to models of realistic materials, this would open the way to perform reliable studies of real materials by performing calculations under the assumptions of one of the transport regimes.

\section*{Data availability} The data that support the findings of this article are openly available \cite{data_zenodo.15273522}.

\acknowledgments

{The authors thank Veljko Jankovi\'c for useful discussions. This research was supported by the Science Fund of the Republic of Serbia, Grant No. 5468, Polaron Mobility in Model Systems and Real Materials - PolMoReMa. The authors acknowledge funding provided by the Institute of Physics Belgrade through a grant from the Ministry of Science, Technological Development, and Innovation of the Republic of Serbia {and support by Serbia Accelerating Innovation and Growth Entrepreneurship (SAIGE) project within the SEED grant for young researchers}. Numerical computations were performed on the PARADOX-IV supercomputing facility at the Scientific Computing Laboratory, National Center of Excellence for the Study of Complex Systems, Institute of Physics Belgrade.}

\clearpage
\pagebreak
\newpage

\setcounter{equation}{0}
\setcounter{figure}{0}
\setcounter{table}{0}
\setcounter{page}{1}
\setcounter{section}{0}
\makeatletter
\renewcommand{\theequation}{S\arabic{equation}}
\renewcommand{\thefigure}{S\arabic{figure}}
\renewcommand{\thetable}{S\arabic{table}}
\renewcommand{\thepage}{S\arabic{page}}
\renewcommand{\thesection}{S\arabic{section}}

\onecolumngrid
\begin{center}
  \textbf{\large Supplemental Material: Identification of the transport regimes of the one-dimensional Holstein model}\\[.2cm]
   Suzana Miladi\'c, Nenad Vukmirovi\'c \\[.1cm]
  {\itshape Institute of Physics Belgrade,
University of Belgrade, Pregrevica 118, 11080 Belgrade, Serbia}
  \\[1cm]
\end{center}

\section{Self-energies in band transport regime}
\noindent In this section of Supplemental Material, we give expressions for self-energies in the band transport regime.

\noindent
The self-energy in the Migdal approximation is given as
\begin{equation}
\label{self_e}
\begin{split}
    \Sigma_k(\omega)=\dfrac{G^2}{N}\sum_q \left[(n_{\mathrm{ph}}+1)G_{k-q}^{(0)}(\omega-\omega_0)+ n_{\mathrm{ph}}G_{k-q}^{(0)}(\omega+\omega_0) \right]\phantom{.},
\end{split}
\end{equation}
where $G_k^{(0)}(\omega)=(\omega-\varepsilon_k+\I0^+)^{-1}$ is the Green's function in the absence of interaction.
By performing the summation in Eq.~\eqref{self_e} one obtains
\begin{equation}
\Sigma(\omega)=G^2(n_{\mathrm{ph}}+1)S(\omega-\omega_0)+G^2n_{\mathrm{ph}}S(\omega+\omega_0)\phantom{.},
\end{equation}
where
\begin{equation}
   S(\omega)=\begin{cases}
            \dfrac{\mathrm{sgn}(\omega)}{\sqrt{\omega^2-4J^2}} \phantom{.}, & \text{if $\vert\dfrac{\omega}{2J}\vert>1$} \\
            \dfrac{-\I}{\sqrt{4J^2-\omega^2}} \phantom{.}, & \text{if $\vert \dfrac{\omega}{2J}\vert<1$}
            \end{cases} \phantom{.}.
\end{equation}
The retarded Green's function is then simply found from the Dyson equation:
\begin{equation}
    G_k^R(\omega)=\dfrac{1}{\omega-\varepsilon_k-\Sigma_k(\omega)} \phantom{.},
\end{equation}
the spectral function is given as
\begin{equation}
A_k(\omega)=-\frac{1}{\pi}\Im G_k^R(\omega)
\end{equation}
and the averages that enter the expression for current-current correlation function read
\begin{equation}
\langle a_k^{\dagger}(t)a_k\rangle = \int_{-\infty}^{\infty}\mathrm{d}\omega \phantom{.} \E^{\I\omega t}A_k(\omega)\E^{-\beta\omega} \phantom{.},
\end{equation}
\begin{equation}
\langle a_k(t)a_k^{\dagger}\rangle=\int_{-\infty}^{\infty}\mathrm{d}\omega \phantom{.}\E^{-\I \omega t}A_k(\omega) \phantom{.}.
\end{equation}

In the case when $\frac{\omega_0}{2J}>1$, we also include the most relevant diagram including two phonon processes. The relevant self-energy is
\begin{equation}
 \Im\Sigma\qty(k,\omega)=
 \frac{g^4}{N^2}n_{\mathrm{ph}}\qty(n_{\mathrm{ph}}+1)
 \sum_{k_1} \Im\frac{1}{\omega-\varepsilon_{k_1}+\mathrm{i}0^+}
 \sum_{k_2}
  \qty[
 \frac{1}{\qty(\varepsilon_{k_2}+\omega_0-\omega)^2}+
 \frac{1}{\qty(\varepsilon_{k_2}-\omega_0-\omega)^2}
 ]
\end{equation}
We then obtain
\begin{equation}\label{eq:si4-wca} 
 \Im \Sigma\qty(\omega) = g^4 n_{\mathrm{ph}} \qty(n_{\mathrm{ph}}+1)
    S_1 \qty[S_2\qty(\omega-\omega_0)+S_2\qty(\omega+\omega_0)]
\end{equation}
where
\begin{equation}
 S_1=\frac{1}{N}\sum_{k}\Im\frac{1}{\omega-\varepsilon_{{k}}+\I 0^+},
\end{equation}
\begin{equation}
 S_2\qty(x)=\frac{1}{N}\sum_{{k}} \frac{1}{\qty(\varepsilon_{{k}}-x)^2}.
\end{equation}
The last two sums can be evaluated analytically and read
\begin{equation}
S_1(\omega)=\begin{cases}
            \dfrac{-1}{\sqrt{4J^2-\omega^2}}\phantom{.}, &\text{if $\vert \dfrac{\omega}{2J}\vert < 1$}\\
            0 \phantom{.}, &\text{otherwise}
        \end{cases}
\end{equation}
\begin{equation}
S_2(x)=\begin{cases}
            -\dfrac{z_1+z_2}{J^2(z_1-z_2)^3}\phantom{.},&\text{if $\dfrac{x}{2J}>1$}\\
            \dfrac{z_1+z_2}{J^2(z_1-z_2)^3}\phantom{.}, &\text{if $\dfrac{x}{2J}<-1$}
        \end{cases}\phantom{.}
\end{equation}
where
\begin{equation}
  z_{1,2}=\dfrac{1}{2}\left(-\dfrac{x}{J}\pm \sqrt{\dfrac{x^2}{J^2}-4} \right).
\end{equation}

\newpage

\section{Self-energy in polaron band transport regime}

The first nonzero term in the retarded self-energy arising from interaction $\tilde{V}$ reads [special case of equations in Ref.~\cite{prb99-104304} when Lang Firsov unitary transformation is used rather than the more general transformation]
\begin{equation}
 \Sigma_k\qty(\omega)=
 \frac{\mathrm{i}}{2\pi}\frac{1}{N^2}
 \sum_q \int \dd \omega_1 G_{k-q}^{\mathrm{R}}\qty(\omega-\omega_1) D^>_{k-q,k,q}\qty(\omega_1)
\end{equation}
with
\begin{equation}
 D^>_{k-q,k,q}\qty(\omega)=
  -\mathrm{i}
  \sum_{R_1S_1R_2S_2} b_{R_1S_1R_2S_2kq}
  \int \dd t \: e^{\mathrm{i}\omega t}
  \qty{e^{a_{R_1R_2S_1S_2}
  \qty[\qty(n_{\mathrm{ph}}+1)
         e^{-\mathrm{i}\omega_0t}+
         n_{\mathrm{ph}} e^{\mathrm{i}\omega_0t} ]}-1},
\end{equation}
\begin{equation}
 b_{R_1S_1R_2S_2kq}=J^2 \delta_{R_1,S_1\pm 1} \delta_{R_2,S_2\pm 1}
 e^{\mathrm{i}\qty[kR_1-\qty(k-q)S_1]}
 e^{\mathrm{i}\qty[\qty(k-q)R_2-kS_2]}.
\end{equation}
Making use of
\begin{equation}
 e^{a\cos\theta}=\sum_{l=-\infty}^{\infty}
 I_l\qty(a)e^{\mathrm{i}l\theta}
\end{equation}
and
\begin{equation}
 \qty(n_{\mathrm{ph}}+1)
         e^{-\mathrm{i}\omega_0t}+
         n_{\mathrm{ph}} e^{\mathrm{i}\omega_0t}
         =2\sqrt{n_{\mathrm{ph}}\qty(n_{\mathrm{ph}}+1)}
         \cos\qty[\omega_0\qty(t+\mathrm{i}\frac{\beta}{2})]
\end{equation}
we arrive at
\begin{equation}
\begin{split}
 \Sigma_k\qty(\omega)&=
 \frac{1}{N^2}\sum_q\sum_{R_1S_1R_2S_2} b_{R_1S_1R_2S_2kq} \\
 &\left\{\qty[I_0\qty(a_{R_1R_2S_1S_2}\cdot 2\sqrt{n_{\mathrm{ph}}\qty(n_{\mathrm{ph}}+1)})-1]G_{k-q}^{\mathrm{R}}\qty(\omega)+\right. \\
 &\left. \sum_{l\ne 0} I_l\qty(a_{R_1R_2S_1S_2}\cdot 2\sqrt{n_{\mathrm{ph}}\qty(n_{\mathrm{ph}}+1)})e^{-\frac{1}{2}l\omega_0\beta}  G_{k-q}^{\mathrm{R}}\qty(\omega+l\omega_0)\right\}.
\end{split}
\end{equation}
In the polaron band transport regime, the Green's function has maxima at energies around $\tilde{\varepsilon}_k$. The dispersion $\tilde{\varepsilon}_k$ is rather flat. Relevant arguments of self-energies are then those at around $\tilde{\varepsilon}_k$. On the other hand, the term $G_{k-q}^{\mathrm{R}}\qty(\omega+l\omega_0)$ has maxima at energies around $\tilde{\varepsilon}_k+l\omega_0$. Hence, it is only the $l=0$ term in previous equation that determines the values of the self-energy at its relevant arguments. By including this term only and introducing the definition
\begin{equation}
S_0\qty(R,\omega,J)=\frac{1}{N}\sum_k\frac{e^{\mathrm{i}kR}}{\omega+2J\cos k+\mathrm{i}0^+}
\end{equation}
we arrive at
\begin{equation}
\begin{split}
\Sigma_k\qty(\omega)=&\sum_{R_1=0,S_1,R_2,S_2} J^2 \delta_{R_1,S_1\pm 1} \delta_{R_2,S_2\pm 1}
e^{-2\qty(\frac{g}{\omega_0})^2\qty(2n_{\mathrm{ph}}+1)}
e^{\mathrm{i}k\qty(R_1-S_2)} \\
& S_0\qty(R_2-S_1,\omega,\tilde{J})
\qty[I_0\qty(2a_{R_1R_2S_1S_2}\sqrt{n_{\mathrm{ph}}\qty(n_{\mathrm{ph}}+1)})-1].
\end{split}
\end{equation}
The sum $S_0\qty(R,\omega,J)$ can be evaluated analytically. To write down this expression we introduce the notation $\tau=\frac{\omega}{2J}$, $z_1=-\tau+\sqrt{\tau^2-1}$, $z_2=-\tau-\sqrt{\tau^2-1}$ and obtain
\begin{equation}
 S_0\qty(R,\omega,J)=
 \begin{cases}
   \frac{z_1^R}{J\qty(z_1-z_2)} &  \tau>-1\:\mathrm{and}\: R\ge 0 \\
   \frac{z_2^R}{J\qty(z_2-z_1)} &  \tau<-1\:\mathrm{and}\: R\ge 0 \\
   \frac{z_1^R}{J\qty(z_1-z_2)}+\frac{1}{Jz_1z_2} & \tau>-1\:\mathrm{and}\: R=-1 \\
   \frac{z_2^R}{J\qty(z_2-z_1)}+\frac{1}{Jz_1z_2} & \tau<-1\:\mathrm{and}\: R=-1 \\
   \frac{z_1^R}{J\qty(z_1-z_2)}+\frac{1}{Jz_1z_2}\qty(\frac{1}{z_1}+\frac{1}{z_2}) & \tau>-1\:\mathrm{and}\: R=-2 \\
   \frac{z_2^R}{J\qty(z_2-z_1)}+\frac{1}{Jz_1z_2}\qty(\frac{1}{z_1}+\frac{1}{z_2}) & \tau<-1\:\mathrm{and}\: R=-2
 \end{cases}
\end{equation}

\newpage

\section{Transport regime crossovers in Fig. \ref{transport_regimes_corrected} for $\mathbf{\omega_0=J/3}$.}

In this section, we explore further into the regime crossovers depicted in Fig. \ref{transport_regimes_corrected} for $\omega_0=J/3$.

\begin{figure}[!h]
    \centering
   \includegraphics[width=0.45\textwidth]{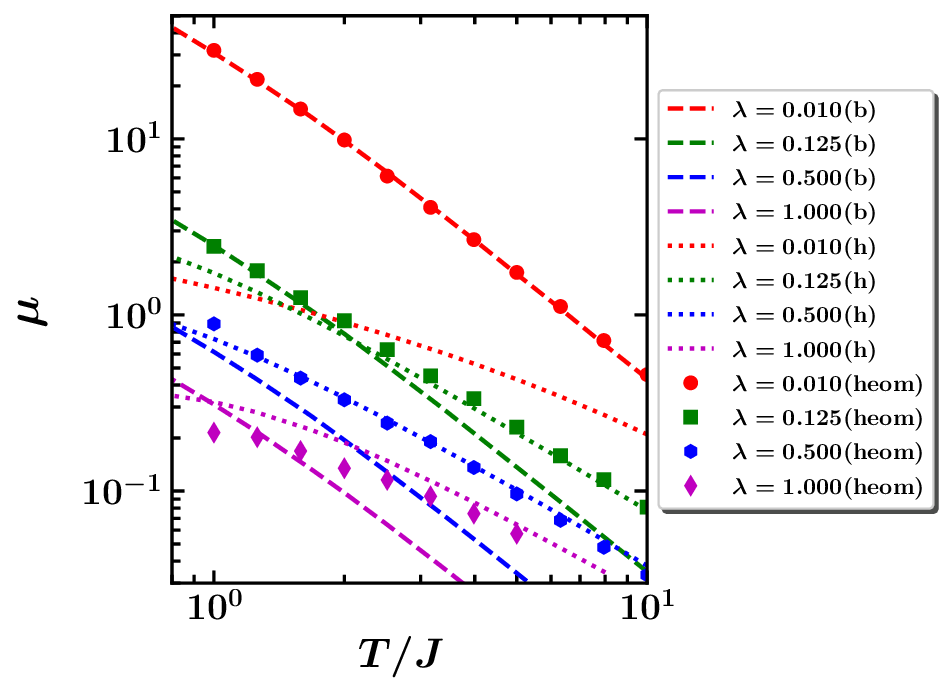}
  \caption{Mobility obtained from formulas for band transport (labeled "b") and hopping transport (labeled "h") compared with numerically exact HEOM data. Numerical data are presented as points. Results are shown for $\omega_0=J/3$.}
  \label{mobility_0.33}
\end{figure}

For the weakest interaction strength where $\lambda=0.010$, the results from Fig. \ref{cmap} imply that both band transport and hopping could be possible in principle at higher temperatures. A closer examination of Fig. \ref{mobility_0.33} shows that the numerically exact data align closely with the band mobility, while the hopping mobility remains below the band mobility curve (see also Fig. \ref{fig4}). This observation confirms that we are indeed dealing with  band transport regime across the entire temperature range.

\begin{figure}[!h]
    \centering
    \includegraphics[width=0.45\textwidth]{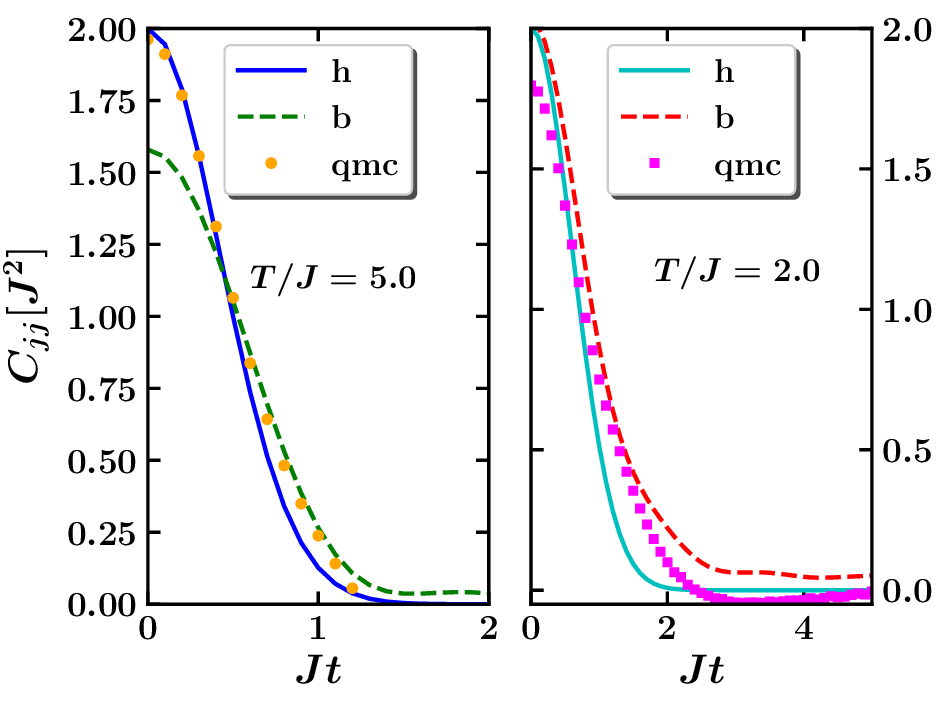}
  \caption{Real time current-current correlation functions for $\omega_0=J/3$, $\lambda=0.125$, and the temperatures $T/J=2.000$ and $T/J=5.000$. Lines labeled with "h" correspond to functions obtained for hopping transport, lines labeled "b" correspond to functions obtained with expression for band transport and QMC results are represented with points.}
  \label{slika_5s1}
\end{figure}

At an interaction strength of $\lambda=0.125$ and temperatures lower than $T/J=1.0$, Fig. \ref{cmap} suggest that band transport is the relevant regime.
However, for temperatures of $T/J=2.0$ and $5.0$, it is unclear whether the correct transport regime is band transport or hopping. Fig. \ref{mobility_0.33} shows mobility data where, in the temperature interval from $T/J=1.0$ to $T/J=10.0$, we are unable to distinguish between band and hopping mechanisms. It is also difficult to distinguish between the two mechanism from real-time correlation functions, as illustrated in Fig. \ref{slika_5s1}, where the QMC data show proximity to both the hopping and band transport correlation function. For this reason, we label this region as an intermediate region between band and hopping transport. Using the results from Fig. \ref{cmap}, we assign the hopping mechanism at largest temperature for this interaction strength.

\begin{figure}[!h]
    \centering
    \includegraphics[width=0.45\textwidth]{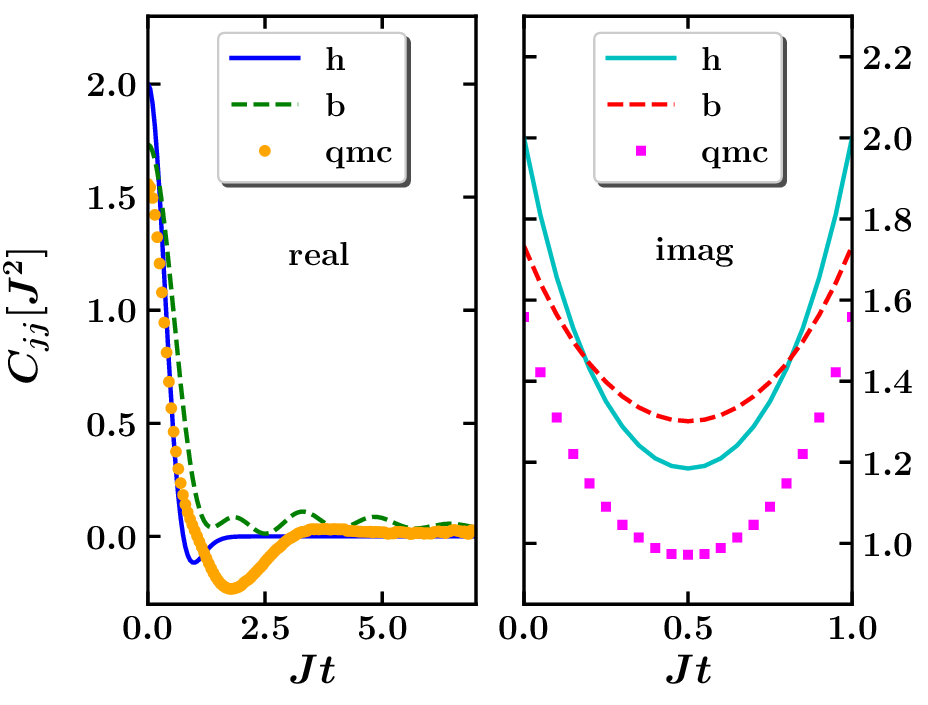}
  \caption{Real and imaginary time current-current correlation functions for  $\omega_0=J/3$, $\lambda=0.500$, and $T/J=1.000$. Lines labeled with "h" correspond to functions obtained for hopping transport, lines labeled "b" correspond to functions obtained with expression for band transport and QMC results are represented with points.}
  \label{slika_5s5}
\end{figure}

At an interaction strength $\lambda=0.500$ from Fig. \ref{cmap} we can deduce that for temperatures greater than $T/J=1.0$ hopping is the dominant mechanism. Fig. \ref{mobility_0.33} illustrates that the numerically exact results align closely with the curve representing hopping mobility across the temperature range of $T/J=1.0$ to $T/J=10.0$. However, upon analyzing the correlation functions, it remains ambiguous whether the appropriate mechanism at $T/J=1.0$ is hopping or band transport, as shown in Fig. \ref{slika_5s5}. At temperatures below $T/J=1.0$, the results from Fig. \ref{cmap} suggest that none of the three regimes is in place. This is depicted in Fig. \ref{transport_regimes_corrected} as a white region.

For an interaction strength of $\lambda=1.000$, as indicated by Figures \ref{cmap} and \ref{mobility_0.33}, it is clear that within the temperature interval from $T/J=1.0$ to $T/J=10.0$, the hopping mechanism is in place. The correlation function data from Fig.~\ref{cmap} suggest that imaginary-time correlation functions correspond most closely to the hopping functions for temperatures below $T/J=1.0$. Fig. \ref{fig3} suggests that the polaron band transport regime can occur only at temperatures below $T/J=0.1$. Consequently, we infer that within the temperature range of $T/J=0.1$ to $T/J=10.0$, hopping transport is dominant for this set of parameters. The same can be concluded for an interaction strength of $\lambda=2.000$.

\newpage
\pagebreak

\section{Transport regime crossovers in Fig. \ref{transport_regimes_corrected} for $\mathbf{\omega_0=3J}$.}

\begin{figure}[!h]
    \centering
    \includegraphics[width=0.45\textwidth]{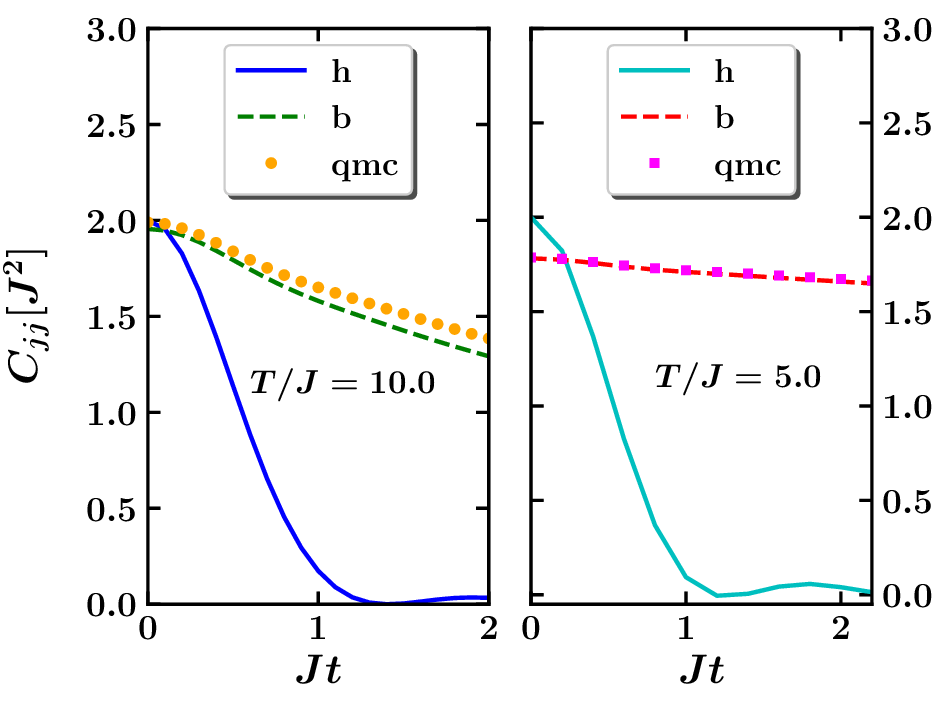}
  \caption{Real time current-current correlation functions for $\omega_0=3J$, $\lambda=0.010$, and the temperatures $T/J=2.000$ and $T/J=10.000$. Lines labeled with "h" correspond to functions obtained for hopping transport, lines labeled "b" correspond to functions obtained with expression for band transport and QMC results are represented with points.}
  \label{slika_5s2}
\end{figure}

Next, we examine Fig. \ref{cmap} in the case $\omega_0=3J$ and $\lambda=0.010$. It is evident that both band transport and hopping meet the criterion within the temperature range of $T/J=1.0$ to $T/J=10.0$. We distinguish between the two through the analysis of real-time current-current correlation functions (see Fig. \ref{slika_5s2}) which imply that within the temperature range $T/J=0.5$ to $T/J=10.0$, the proper regime is conventional band transport. This conclusion can be safely extended to lower temperatures based on physical arguments that lowering the temperature for weak interaction reduces the electron-phonon scattering and therefore preserves the band transport mechanism.

\begin{figure}[!h]
    \centering
   \includegraphics[width=0.45\textwidth]{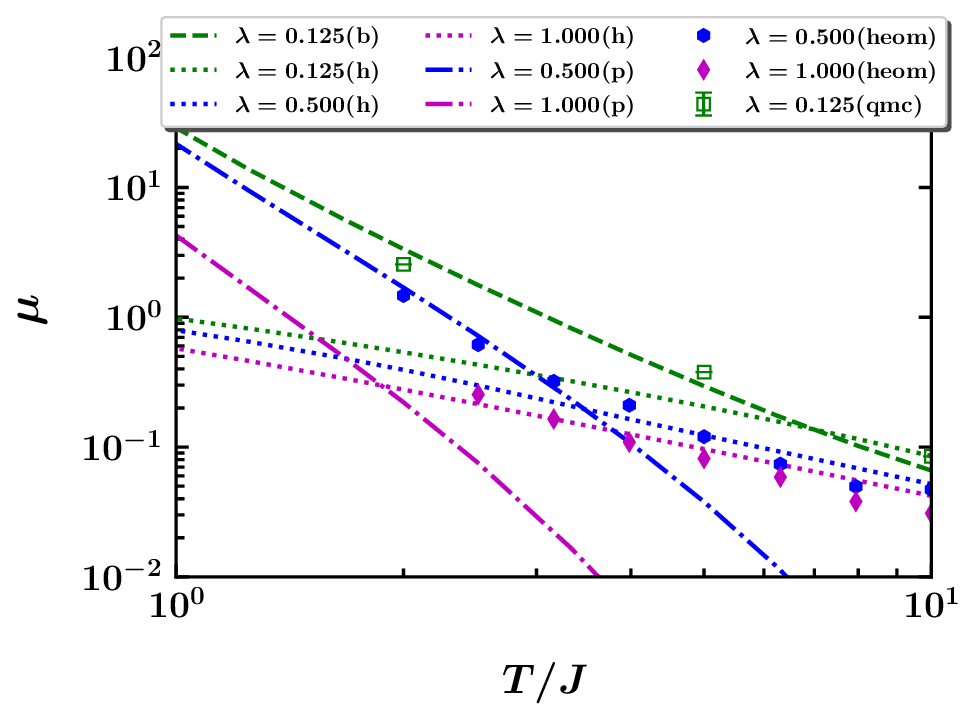}
  \caption{Mobility obtained from formulas for band transport (labeled "b"), hopping transport (labeled "h") and polaron band transport (labeled "p") compared with numerically exact heom data. Numerical data are presented as points. Results are shown for $\omega_0=3J$.}
  \label{mobility_3.0_s}
\end{figure}

\begin{figure}[!h]
    \centering
    \includegraphics[width=0.45\textwidth]{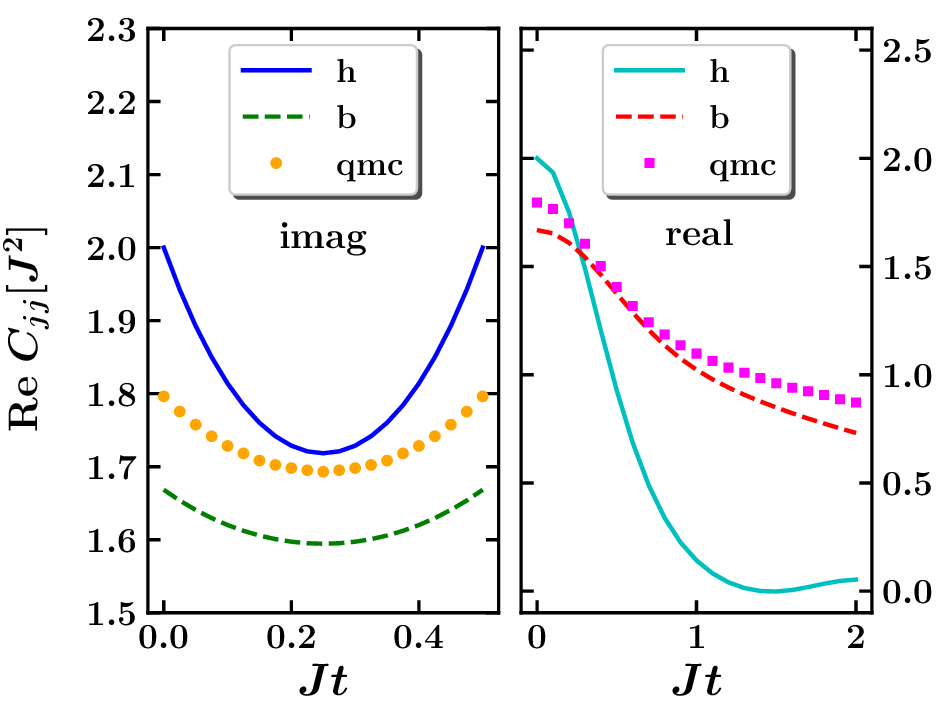}
  \caption{Imaginary- and real-time current-current correlation functions shown for specified parameters $\omega_0=3J$, $\lambda=0.125$, $T/J=2.000$. Lines labeled with "h" correspond to functions obtained with Eq.~\ref{Cjj_slA} for hopping transport, lines labeled "b" correspond to functions obtained with expression for band transport and QMC results are represented with points.}
  \label{slika_5b}
\end{figure}

\begin{figure}[!h]
    \centering
    \includegraphics[width=0.45\textwidth]{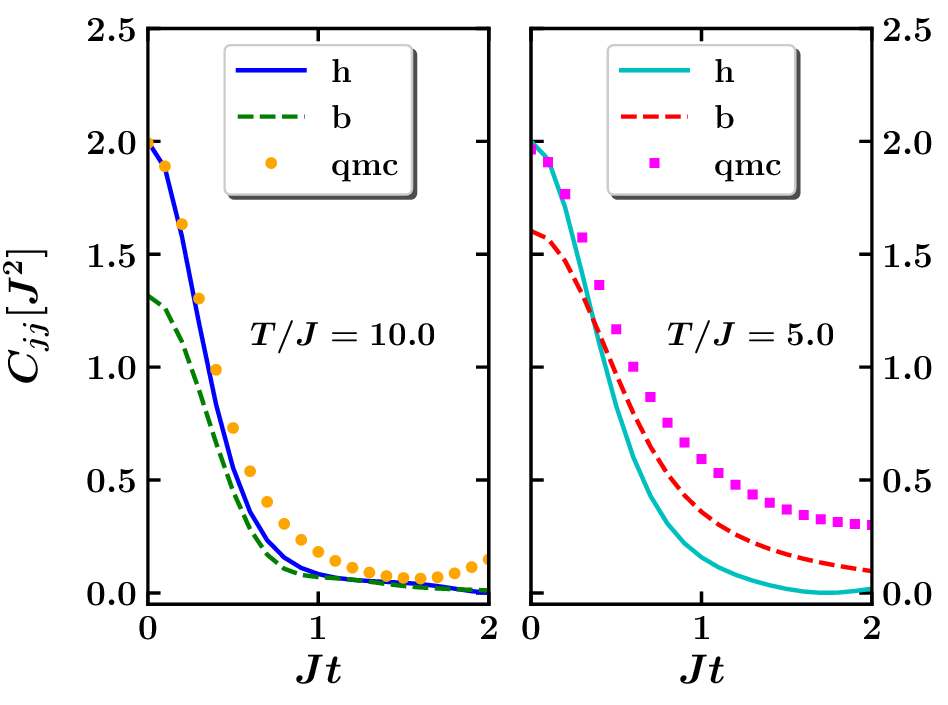}
  \caption{Real-time current-current correlation functions for $\omega_0=3J$, $\lambda=0.125$, and the temperatures $T/J=5.000$ and $T/J=10.000$. Lines labeled with "h" correspond to functions obtained for hopping transport, lines labeled "b" correspond to functions obtained with expression for band transport and QMC results are represented with points.}
  \label{slika_5s3}
\end{figure}

When considering the subsequent interaction strength of $\lambda=0.125$, as shown in Fig. \ref{cmap}, the imaginary-time criterion is met for band transport between $T/J=0.2$ and $T/J=5$, and for hopping in the temperature range from $T/J=1$ to $T/J=10$. The limited mobility data (refer to Fig. \ref{mobility_3.0_s}) indicate that for temperatures lower than $T/J=5$, the appropriate regime is band transport. This can also be seen from the real time correlation function shown in Fig. \ref{slika_5b} for $T/J=2$, which demonstrates how real-time QMC data more closely follow the curve for band transport. We continue to explore the real-time current-current correlation functions. As illustrated in Fig. \ref{slika_5s3}, at a temperature of $T/J=5$, the system is in the intermediate regime between band transport and hopping. However, at the higher temperature of $T/J=10.0$, it is evident from Fig. \ref{slika_5s3} and \ref{cmap} that hopping is the relevant regime.

\begin{figure}[!h]
    \centering
    \includegraphics[width=0.45\textwidth]{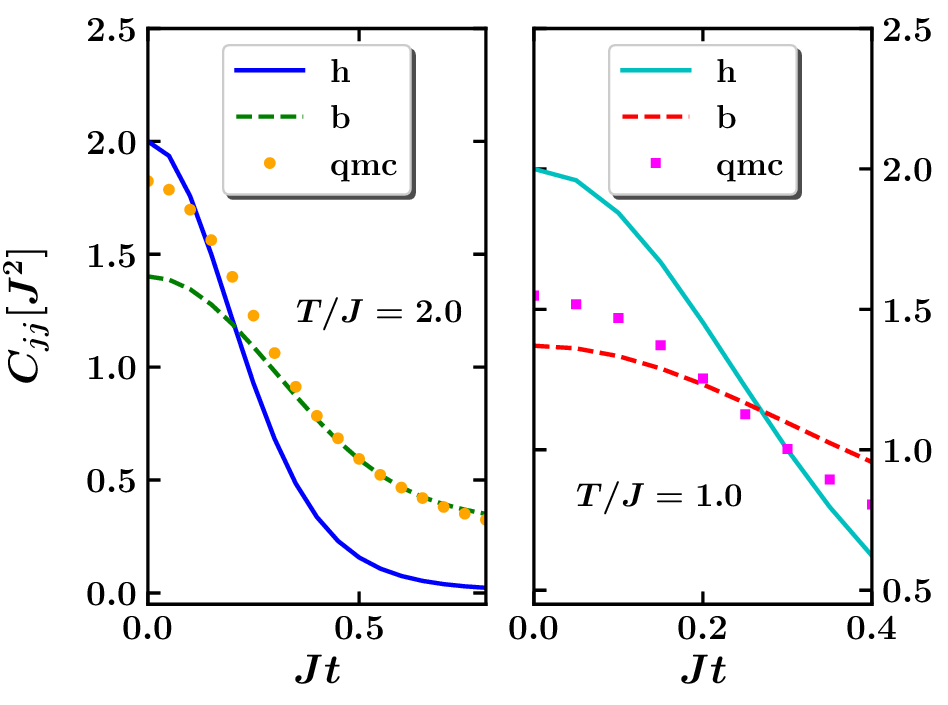}
  \caption{Real-time current-current correlation functions for $\omega_0=3J$, $\lambda=0.500$, and the temperatures $T/J=1.000$ and $T/J=2.000$. Lines labeled with "h" correspond to functions obtained for hopping transport, lines labeled "b" correspond to functions obtained with expression for band transport and QMC results are represented with points.}
  \label{slika_5s4}
\end{figure}

Next, we examine the results for the interaction strength $\lambda=0.500$. We can regard this interaction as moderately strong, suggesting the potential for polaron band transport to occur. As illustrated in Fig. \ref{cmap}, the conditions for polaron band transport are met within the temperature range of $T/J=0.1$ to $T/J=2$. Furthermore, band transport might be possible from $T/J=0.5$ to $T/J=2$, and hopping could occur from $T/J=1$ to $T/J=10$. We exclude the likelihood of conventional band transport at any temperature upon analyzing real time current-current correlation functions. Fig. \ref{slika_5s4} supports this argument as it shows that numerically exact QMC data are equally far from the band transport and hopping regime. Fortunately, there is a limited set of numerically exact HEOM data in this range of parameters, which effectively demonstrate the shift from polaron band transport to hopping, as depicted in Fig. \ref{mobility_3.0_s} for mobility. This transition aligns with the point where the mobility curves for polaron band transport and hopping intersect.

For the two strongest interactions, $\lambda=1.000$ and $\lambda=2.000$, we lack some numerical data at the lowest temperatures. Therefore, our deductions regarding transport regime transitions will be based on previous conclusions. From Fig. \ref{cmap}, for an interaction strength of $\lambda=1.000$,  the criterion for polaron band transport is met within the temperature range of $T/J=0.1$ to $T/J=1.0$, whereas hopping might be possible from $T/J=0.5$ up to $T/J=10$. We dismiss the possibility of conventional band transport for the same reasons as for the previous interaction strength. As depicted in figures \ref{fig3} and \ref{mobility_3.0_s}, the curves for polaron band mobility and hopping mobility intersect around the temperature $T/J=2$. Thus, we deduce that the appropriate transport mechanism is polaron band transport up to $T/J=2$, while for temperatures exceeding $T/J=2$, the hopping mechanism prevails. In a similar manner, when the interaction is $\lambda=2.000$, the point where the two corresponding curves intersect occurs at approximately $T/J=1.0$. Thus, we conclude that the appropriate transport mechanism is polaron band transport for temperatures lower than $T/J=1$, transitioning to hopping transport for temperatures beginning at $T/J=1.0$ and above.

\section{Details on QMC calculations of imaginary-time current-current correlation functions.}

\begin{figure}[!h]
    \centering
    \includegraphics[width=0.45\textwidth]{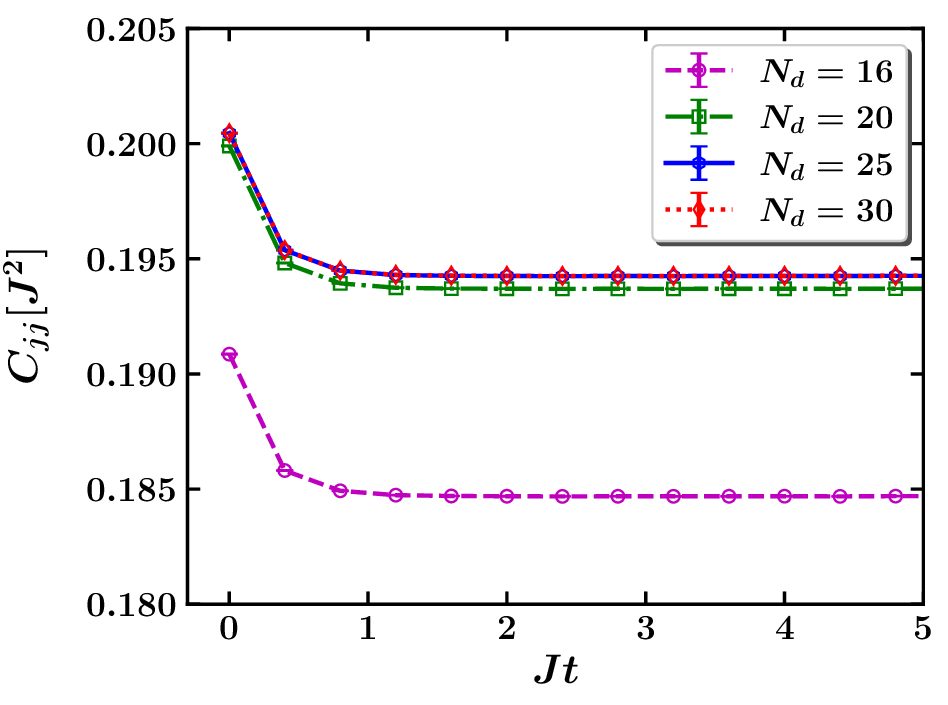}
  \caption{Imaginary-time current-current correlation function calculated with different number of sites $N_d$. The results are shown for the interaction strength $\lambda=0.010$, the temperature $T/J=0.1$ and the phonon frequency $\omega_0=3J$.}
  \label{slika_5s6}
\end{figure}

In this section we present details on the choice of simulation parameters in QMC calculations of imaginary-time current-current correlation functions in present work. In Fig. \ref{slika_5s6} the effect of limited system size (number of sites $N_d$) on result of QMC calculations is shown for the interaction strength $\lambda=0.010$, the temperature $T/J=0.1$ and the phonon frequency $\omega_0=3J$. We can see that the number of sites needed to reach the thermodynamic limit is 25 in that case.

\begin{figure}[!h]
    \centering
    \includegraphics[width=0.95\textwidth, trim=2cm 11cm 2.7cm 2cm, clip]{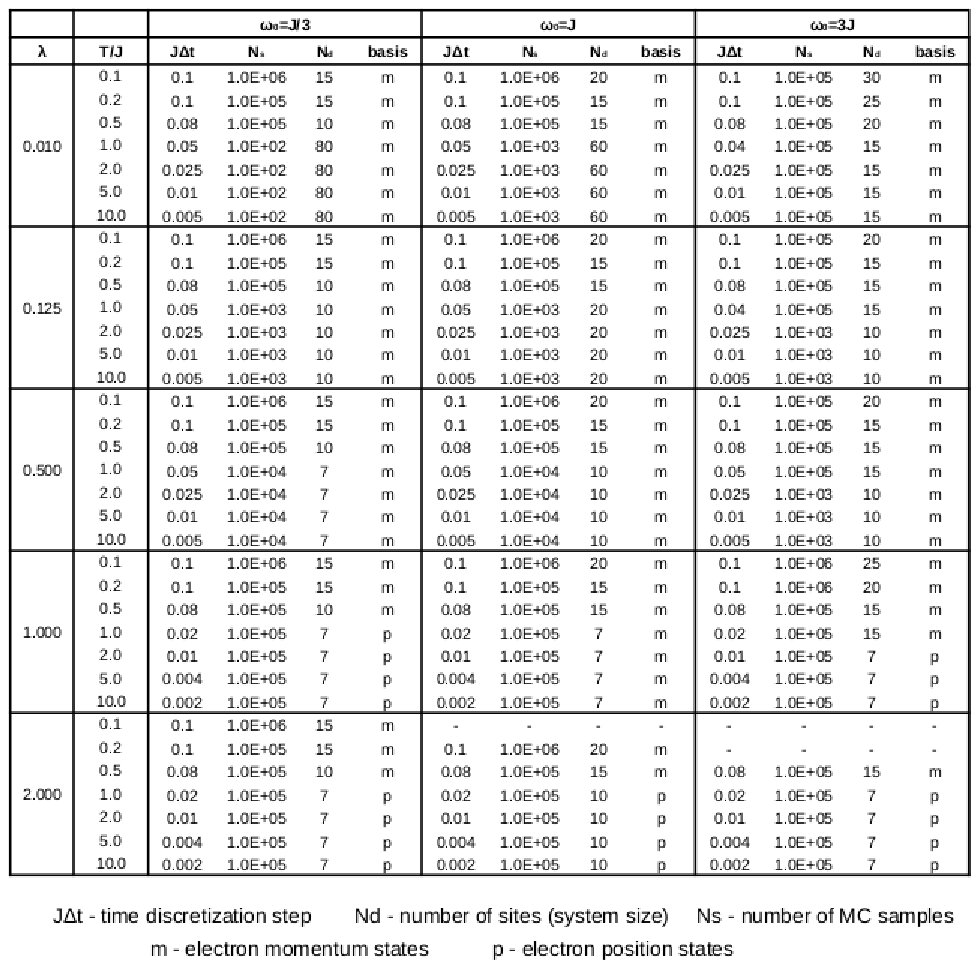}
  \caption{The table with values of simulation parameters used in QMC calculations of imaginary-time current-current correlation functions.}
  \label{slika_5s7}
\end{figure}

In Fig. \ref{slika_5s7} a table is provided with the values used in our calculations for the time discretization step $J\Delta t$, the number of MC samples $N_s$ and the system size (the number of sites) $N_d$ for each specified temperature $T/J$, interaction strength $\lambda$ and phonon frequency $\omega_0$. The table also contains information whether the calculations were completed using the momentum (m) or the electron position (p) basis.

\end{document}